\documentclass[manuscript,screen]{acmart}

\usepackage{hyperref}
\usepackage{csquotes}
\usepackage{enumitem}
\usepackage{tikz}
\usetikzlibrary{positioning, arrows.meta}
\usepackage{tikz}
\usetikzlibrary{shapes.geometric, arrows, fit}
\usepackage{subcaption}
\usepackage{tcolorbox}
\usepackage{xcolor}

\definecolor{primaryBlue}{HTML}{77AADD} 

\tcbset{
  interviewquote/.style={
    colback=primaryBlue!5!white,
    colframe=primaryBlue,
    boxrule=0.25mm,
    arc=2mm,
    left=2mm,
    right=2mm,
    top=1mm,
    bottom=1mm
  }
}

\definecolor{ColorPink}{HTML}{FF8DAE}
\definecolor{ColorOrange}{HTML}{EE8866} 
\definecolor{ColorBlue}{HTML}{77AADD} 
\definecolor{ColorLightblue}{HTML}{99ddff}
\definecolor{ColorGreen}{HTML}{44bb99} 

\setcopyright{rightsretained}
\copyrightyear{2026}
\acmYear{2026}
\acmDOI{XXXXXXX.XXXXXXX}
\author{Shalini  Chakraborty}
\email{shalini.chakraborty@uni-bayreuth.de}
\orcid{0000-0002-9466-3766}
\affiliation{%
  \institution{University of Bayreuth}
  \city{Bayreuth}
  \country{Germany}
}
\author{Marvin Wyrich}
\email{wyrich@cs.uni-saarland.de}
\orcid{0000-0001-8506-3294}
\affiliation{%
  \institution{Saarland University}
  \city{Saarbr\"ucken}
  \country{Germany}
}
\author{Sven Apel}
\email{apel@cs.uni-saarland.de}
\orcid{0000-0003-3687-2233}
\affiliation{%
  \institution{Saarland University}
  \city{Saarbr\"ucken}
  \country{Germany}
}
\author{Sebastian Baltes}
\email{s.baltes@se-uhd.de}
\orcid{0000-0002-2442-7522}
\affiliation{%
  \institution{Heidelberg University}
  \city{Heidelberg}
  \country{Germany}
}

\begin{document}

\title{Motivations and Barriers to Communicating Software Engineering Research: Insights from Early Career Researchers}
\renewcommand{\shorttitle}{Science Communication in SE}

\begin{abstract}
Science communication is increasingly becoming a part of modern research careers, involving researchers to disseminate knowledge, engage broader communities, and increase the societal impact of their work. Despite its growing importance, little is known about how early-career software engineering researchers perceive and navigate science communication in practice. In this paper, we investigate how PhD students in software engineering experience science communication. We conducted semi-structured interviews with 18 doctoral candidates from diverse international backgrounds. Using thematic analysis, we examine three interconnected dimensions: motivations, communication channels, and barriers.
Our findings reveal a strong tension between aspiration and practice. Participants were highly motivated to engage in science communication due to opportunities for collaboration, professional recognition, broader impact, and advocacy for themselves and their research. However, translating these motivations into action was frequently constrained by social anxiety, uncertainty regarding appropriate audiences and communication venues, limited feedback mechanisms, insufficient institutional guidance, and challenges associated with navigating an increasingly fragmented communication landscape. Our findings highlight the need for tailored, balanced support systems that empower software engineering PhD students to engage in science communication effectively and confidently across diverse cultural and institutional environments.
We outline practical implications that offer initial guidance for addressing these challenges in future work.
\end{abstract}

\ccsdesc[300]{Social and professional topics~Socio-technical systems}

\keywords{Science Communication, Software Engineering, Collaboration}

\maketitle

\section{Introduction}
\label{sec:introduction}
Science communication plays a crucial role in making research accessible, relevant, and impactful beyond the boundaries of academic publications~\cite{burns2003science,bucchi2016science}. In software engineering (SE), this is particularly significant, as research outcomes often have direct implications for industry practice, education, and policy~\cite{DBLP:journals/software/DybaKJ05,DBLP:conf/icse/SjobergDJ07}. Despite this relevance, SE researchers vary considerably in how they engage in science communication and in the extent to which they communicate their work outside traditional peer-reviewed venues. 
While some researchers actively pursue outreach and dissemination through blogs, social media, or open-source projects, others remain hesitant or rely exclusively on academic publishing as their primary mode of communication.
%

Prior research has shown that engagement in science communication is influenced by factors such as career stage, institutional expectations, linguistic and cultural background, personal motivation, and gender~\cite{cerrato2018public,poliakoff2007factors,hamlyn2015factors,rodriguez2021perceived}. However, little is known about how these dynamics specifically shape the behavior of SE researchers, a community whose work is closely tied to industrial practice and whose relevance to practice has long been a recurring concern~\cite{garousi2020practical}.
%
SE research often needs to be communicated effectively to a wide range of stakeholders, including academics, industry practitioners, open-source communities, and policymakers. The field’s rapid evolution, particularly with the rise of artificial intelligence and generative technologies, further underscores the importance of clear and accessible communication practices. Moreover, SE researchers seem to employ a diverse set of dissemination channels, ranging from traditional publications to collaborative platforms such as GitHub and professional social media networks, offering a rich landscape for examining how research is shared, interpreted, and utilized. The community’s strong emphasis on transparency, reproducibility, and evidence-based practice makes SE a compelling context for exploring the broader challenges of science communication.

Early-career researchers, particularly PhD students and recent graduates, represent a new generation of academics whose science communication practices and struggles warrant closer investigation. Although engaging in science communication can offer critical advantages such as enhancing visibility, improving research impact, and fostering collaboration, the pathways toward effective engagement are not always clear for novice academics~\cite{peters2013gap,besley2019strategic}. Beyond individual factors such as confidence or personal interest, elements such as the dissertation topic, supervision and mentorship, institutional recognition, and access to training programs may play a significant role in shaping their willingness and ability to engage~\cite{mason2022communicating}. Economic and linguistic constraints may further limit opportunities, while a lack of clarity regarding target audiences and communication channels can exacerbate uncertainty. Such barriers can result in disengagement or demotivation, particularly in environments where science communication is undervalued or unsupported.
%

Understanding these challenges is crucial, since establishing effective communication practices early in an academic career can yield long-term benefits for both individual researchers and the broader societal impact of research. This is especially relevant in SE, where academia and industry are tightly interconnected, and where the communication of empirical findings can facilitate knowledge transfer, improve software development practices, and strengthen collaboration between academia and industry~\cite{mikkonen2018continuous}.

\noindent
\\
\textbf{In this paper, we focus on understanding the perspectives of PhD students on science communication within SE, addressing the following research questions:}
\begin{itemize}

    \item\textbf{RQ1:} What motivates early-career SE researchers to share their research?
   \item\textbf{RQ2:} Which channels do early-career SE researchers use to communicate SE research?
    \item\textbf{RQ3:} What barriers do early-career SE researchers face when communicating SE research?
\end{itemize}

To answer these questions, we conducted a qualitative interview study with 18 participants (including two pilot interviews) who are PhD students at various stages of their doctoral studies and based in different regions of the world. 
The results of our study yield three thematic dimensions that capture how early-career SE researchers think about science communication in terms of \textit{motivations}, \textit{barriers}, and \textit{channels}. Each dimension is supported by multiple themes that reflect how PhD students perceive the value of communication, the strategies or platforms they use to share their work and the challenges they encounter. Taken together, these themes provide a nuanced view of science communication as both an aspiration and a struggle, shaped by enthusiasm for outreach as well as constraints related to confidence, institutional support, and contextual limitations. Rather than proposing a formal framework, our analysis highlights recurring patterns and derives practical implications that can guide the development of targeted interventions and support mechanisms in the SE research community.

The remainder of this paper is structured as follows: In Section~\ref{sec:relatedwork} we summarize related work on science communication in general and in SE. In Section~\ref{sec:method} we describe our methodology, followed by results in Section~\ref{sec:results} and a discussion in Section~\ref{sec:discussions}. We conclude the paper in Section~\ref{sec:conclusion}.

\section{Background and Related Work}
\label{sec:relatedwork}
\subsection{Science Communication: Definitions and Dimensions}
Science communication has broadly been defined as the use of appropriate skills, media, activities, and dialogue to generate responses such as \textit{awareness}, \textit{enjoyment}, \textit{interest}, \textit{opinion-forming}, and \textit{understanding} of science~\cite{burns2003science}. Early perspectives frequently conceptualized communication as a one-way transfer of knowledge from experts to the public, commonly referred to as the \textit{deficit model}. Over time, this perspective has evolved toward dialogic and participatory approaches that emphasize interaction, reflexivity, and mutual engagement between researchers and audiences~\cite{trench2008towards,bauer2009evolution}. Rather than treating communication as a final dissemination stage following research production, these approaches increasingly view communication as an ongoing process embedded within scientific practice itself.
%
Recent studies further argue that science communication extends beyond dissemination and constitutes a broader socio-technical activity involving multiple actors, communication environments, and institutional contexts~\cite{besley2019strategic,davies2016science}. Researchers are increasingly expected not only to communicate scientific findings but also to engage with stakeholders, policymakers, practitioners, and broader communities. As a result, science communication has become intertwined with questions of societal impact, researcher visibility, and public engagement~\cite{national2017communicating,mason2022communicating}.

Existing work additionally highlights that communication practices are shaped by researchers' motivations, institutional structures, communication channels, and disciplinary cultures~\cite{poliakoff2007factors,besley2019strategic}. However, communication practices may vary substantially across domains and communities. Emerging work increasingly argues against universal approaches to science communication and instead emphasizes context-dependent understandings of what communication practices work, for whom, and under which conditions~\cite{Achiam2025Terroir}. Despite these developments, limited work has examined how science communication is perceived and experienced within software engineering communities, particularly among early-career researchers who may encounter distinct motivations and barriers when engaging with diverse stakeholder groups.

\subsection{Science Communication in Technical Disciplines}
While much research on science communication has focused on disciplines such as the life sciences, physics, and climate science, emerging studies are beginning to address communication practices in more technical and data-intensive domains. For example, \citeauthor{eysenbach2009infodemiology}~\cite{eysenbach2009infodemiology} introduced the concept of \textit{infodemiology} and \textit{infoveillance} as frameworks for analyzing how health-related information is searched for and communicated online, illustrating the intersection of informatics and public communication. More recently, \citeauthor{schoning2025health}~\cite{schoning2025health} examined how health informatics researchers frame their findings for non-expert audiences, identifying tensions between precision and accessibility. Similarly, \citeauthor{DBLP:journals/corr/abs-2507-10559}~\cite{DBLP:journals/corr/abs-2507-10559} investigated natural language processing (NLP) researchers’ strategies for engaging the public, revealing how experts balance outreach efforts with concerns about oversimplification or misinterpretation. Collectively, these studies highlight how researchers in technical fields navigate the trade-off between maintaining precision and making their work accessible to non-expert audiences.

\subsection{Science Communication in Software Engineering}
Within software engineering, discussions around science communication remain comparatively recent but are increasingly gaining attention. \citeauthor{DBLP:journals/cacm/WyrichTBA25}~\cite{DBLP:journals/cacm/WyrichTBA25} argue that software engineering researchers have often acted as \enquote{silent scientists}, producing research with potential societal and industrial relevance while implicitly assuming that publication alone is sufficient for impact. Their work challenges this assumption and emphasizes that research impact frequently requires active dissemination through communication channels tailored to relevant audiences~\cite{DBLP:journals/cacm/WyrichTBA25}.

Emerging evidence suggests that such communication practices are already beginning to appear within software engineering communities. For example, recent work has examined the use of social media platforms such as LinkedIn for disseminating software engineering research and highlighted both opportunities for visibility and persistent barriers related to engagement and audience reach~\cite{DBLP:conf/icse/WyrichB24,DBLP:conf/sbes/GarciaHSB25}. Educational initiatives have additionally proposed integrating science communication training into computer science and software engineering curricula to improve researchers' and students' abilities to communicate with broader audiences~\cite{DBLP:conf/icse/WyrichW23}. Beyond social media, researchers have also begun experimenting with alternative dissemination mechanisms, including blogs, podcasts, public-facing research summaries, and specific practitioner-oriented communication formats designed to increase accessibility of software engineering knowledge~\cite{DBLP:journals/corr/abs-2507-10559,DBLP:journals/software/WilsonAHJ24}.

\subsection{Research Gap}
There is a consensus that science communication is important for making research accessible to specific target audiences.
In software engineering, too, we need to ensure that such communication goes beyond traditional publications and conference presentations.
However, much of the existing work focuses on describing individual initiatives or stops at advocating for stronger dissemination practices. To the best of our knowledge, no empirical work examines how SE researchers themselves perceive, experience, and navigate science communication activities in practice, particularly during early stages of academic careers. Consequently, while the software engineering community increasingly recognizes the importance of science communication, our understanding of the motivations, communication channels, and experienced barriers remains limited.
Answering the research questions outlined in Section~\ref{sec:introduction} is intended to fill this gap, and in the following Section~\ref{sec:method}, we describe our approach to addressing them.

\section{Method}
\label{sec:method}

SE is a global and diverse discipline, encompassing researchers from a wide range of professional, cultural, linguistic, and seniority backgrounds. In this study, we focus on PhD students at different stages of their doctoral journey. As the next generation of SE scholars and practitioners, PhD students offer a unique vantage point on emerging science communication practices and norms, while simultaneously navigating the opportunities and challenges faced by early-career researchers within both academic and industrial ecosystems.

To capture these perspectives, we conducted semi-structured interviews with PhD students recruited from multiple universities across different geographic regions. 
Ethical approval for this study was obtained from the first author’s host institution prior to data collection. The approval specified procedures for participant recruitment, data handling, and data protection. Before each interview, the participants received an informed consent form describing the study’s purpose, its voluntary nature, their right to withdraw at any time, and the intended use of the collected data for research and publication. All interview data were anonymized and securely stored to ensure participant confidentiality.

The interview guide consists of an introductory section, a \emph{warm-up section} to make the interviewee feel comfortable by simply telling us about their current research, followed by questions on participants' motivations for science communication, the channels they use, and the barriers they encounter, and concludes with two reflective questions on the perceived need for further outreach and the most significant challenge they face when promoting their research.
The complete interview guide as well as data analysis materials are provided as a replication package archived on Zenodo~\cite{chakraborty2026replication}.


\subsection{Pilot Study} 
Before beginning the main study, we conducted two pilot interviews to refine the interview guide and determine the appropriate duration for each session. The first and second authors each selected one PhD student from their personal networks to participate. Both authors were present during the pilot interviews, and consent forms were sent to the participants in advance, and we asked for permission to record the sessions. Although we initially planned for one-hour interviews, both pilots concluded within 30–40 minutes, leading us to set the final interview duration at 40 minutes. Based on insights from the pilot data, we added two new questions to the interview guide (both were potential follow-up questions to a question about target audiences). 
These additions are noted in the guide, which is included in the supplemental materials.

\subsection{Data Collection}

We conducted semi-structured interviews with PhD students in SE, targeting participants through the Doctoral Symposium tracks of two leading conferences: the International Conference on Software Engineering (ICSE) and the International Conference on the Foundations of Software Engineering (FSE). We contacted all authors accepted to these tracks over the past two years (2024 and 2025) via email, selecting this timeframe to ensure that participants were likely still enrolled as PhD students.
ICSE and FSE were chosen for their global reputation and prestige within the software engineering community, as well as their Doctoral Symposium tracks, which provide a practical and systematic method to identify and recruit early-career researchers. While this approach limits the sample to students whose submissions were accepted, we contend that leveraging these established venues known for their relatively high acceptance rates helps mitigate selection bias. A more detailed discussion of this limitation and our rationale is provided in Section~\ref{sec:validity}.

In total, we contacted 35 participants from ICSE'24, 12 from FSE'24, 42 from ICSE'25, and 19 from FSE'25. From these, we successfully scheduled and conducted interviews with 16 participants. Including two pilot interviews, our final dataset comprises 18 interviews.
Interviews were conducted by the first two authors. In 13 of the 16 formal interviews, both authors were present; in the remaining three, one author conducted the session due to scheduling constraints. After each interview, the two authors engaged in reflective discussions and began to observe recurring patterns related to how participants experience science communication, and diminishing novelty in the data after approximately seven interviews, indicating emerging information saturation. We nevertheless proceeded with all scheduled interviews to strengthen the robustness and confidence in our findings.

\subsection{Data Analysis}
We analyzed the interview data using hybrid deductive--inductive thematic analysis~\cite{braun2006using}, relying on the full set of interview transcripts as the primary data source.
%
To establish a shared analytic framework, the first and second authors jointly coded the first interview. This initial coding led to the identification of four primary categories: \textit{demographics}, \textit{communication channels}, \textit{motivations}, and \textit{barriers}. The second and third interviews were then coded independently by both authors, followed by discussions to reconcile discrepancies. After coding these three interviews, a second level of categorization was introduced under the primary categories, referred to as \textit{themes}. For example: \textit{barriers} $\rightarrow$ lack of time / lack of feedback, \textit{motivations} $\rightarrow$ recognition / recruitment. At this stage, the authors also determined that coding a single interview typically required 40–45 minutes.

To strengthen the analysis, both authors independently coded the next five interviews and then compared their coding to discuss differences, clarify interpretations, and reach consensus on the code definitions.
With eight interviews completed (one collaboratively, seven independently), sufficient material was available to assess consistency. The authors met multiple times to evaluate two aspects: (1) assignment of primary categories, and (2) identification and labeling of themes. 
At this stage, the authors had reached an agreement on the assignment of primary categories. However, there were still minor discrepancies in the labeling of themes, as each author had used slightly different wording to describe similar concepts. To manage these differences efficiently and maintain consistency across the dataset, the authors decided to code the remaining interviews independently. The thematic mapping and consolidation of similar themes were then performed afterward, ensuring that all perspectives were systematically reconciled into a coherent framework.

Once all interviews were coded, the authors conducted a systematic review of the coding results to merge similar themes and reconcile any remaining discrepancies. For each primary category, the authors compared their independently coded data line by line, identifying overlaps, refining labels, and grouping related coded data into coherent themes. Divergent interpretations were discussed in detail, and consensus was reached through iterative negotiation, ensuring that each theme accurately reflected the perspectives expressed by participants. Sub-themes were also identified to capture nuanced variations within broader themes, providing a more granular representation of the data. This collaborative process resulted in the development of three comprehensive thematic maps corresponding to \textit{motivations} (Fig~\ref{fig:map-motivations}), \textit{barriers} (Fig~\ref{fig:map-barriers}), and \textit{communication channels}. Each thematic map integrates the voices of all 18 interviewees, illustrating both shared experiences and individual variations. 

\subsubsection{Member Checking}
Given the interpretive nature of our qualitative analysis, we applied member checking procedure to enhance the credibility and trustworthiness of our findings. After completing the initial analysis of the interview data, we contacted all 18 interview participants via email and invited them to review the thematic maps of motivation and barriers. 

\begin{figure}[t]
\centering
\begin{tikzpicture}[
    node distance=0.55cm,
    box/.style={
        rectangle,
        rounded corners,
        draw=black!65,
        thick,
        align=center,
        text width=7.2cm,
        minimum height=1.05cm,
        font=\small
    },
    data/.style={box, fill=ColorBlue!30},
    analysis/.style={box, fill=ColorOrange!20},
    validation/.style={box, fill=ColorGreen!30},
    arrow/.style={-{Latex[length=2.5mm]}, thick}
]

\node[data] (recruit) {
\textbf{Participant Recruitment}\\
ICSE'24: 35, FSE'24: 12\\
ICSE'25: 42, FSE'25: 19\\
Total contacted: 108
};

\node[data, below=of recruit] (interviews) {
\textbf{Interview Data Collection}\\
16 formal semi-structured interviews + 2 pilot interviews\\
Final dataset: 18 interview recordings
};

\node[data, below=of interviews] (prep) {
\textbf{Data Preparation}\\
Transcription using Microsoft Teams and Adobe Podcasts\\
Full transcripts used as primary data source
};

\node[analysis, below=of prep] (joint) {
\textbf{Initial Collaborative Coding}\\
First interview jointly coded by the first two authors\\
Initial categories identified: demographics, channels, motivations, barriers
};

\node[analysis, below=of joint] (independent) {
\textbf{Independent Coding and Comparison}\\
Interviews 2--8 independently coded by two authors\\
Coding compared for category assignment and theme labeling
};

\node[analysis, below=of independent] (remaining) {
\textbf{Remaining Interviews and Theme Consolidation}\\
Remaining interviews coded independently\\
Similar themes merged, labels refined, and sub-themes identified
};

\node[validation, below=of remaining] (member) {
\textbf{Member Checking}\\
Preliminary thematic maps shared with all 18 participants\\
7 participants responded; feedback incorporated into final interpretation
};

\node[validation, below=of member] (outputs) {
\textbf{Final Thematic Outputs}\\
Thematic maps for motivations, barriers, and communication channels
};

\draw[arrow] (recruit) -- (interviews);
\draw[arrow] (interviews) -- (prep);
\draw[arrow] (prep) -- (joint);
\draw[arrow] (joint) -- (independent);
\draw[arrow] (independent) -- (remaining);
\draw[arrow] (remaining) -- (member);
\draw[arrow] (member) -- (outputs);

\end{tikzpicture}
\caption{Overview of the data collection and qualitative analysis workflow, including recruitment, interview collection, transcription, iterative coding, reconciliation, member checking, and development of the final thematic maps.}
\Description{A vertical flowchart of eight stages: participant recruitment (108 contacted across ICSE'24/'25 and FSE'24/'25), interview data collection (16 formal plus 2 pilot interviews), data preparation (transcription), initial collaborative coding, independent coding and comparison, consolidation of remaining interviews and themes, member checking with participants, and the final thematic outputs.}
\label{fig:data-collection-analysis}
\end{figure}
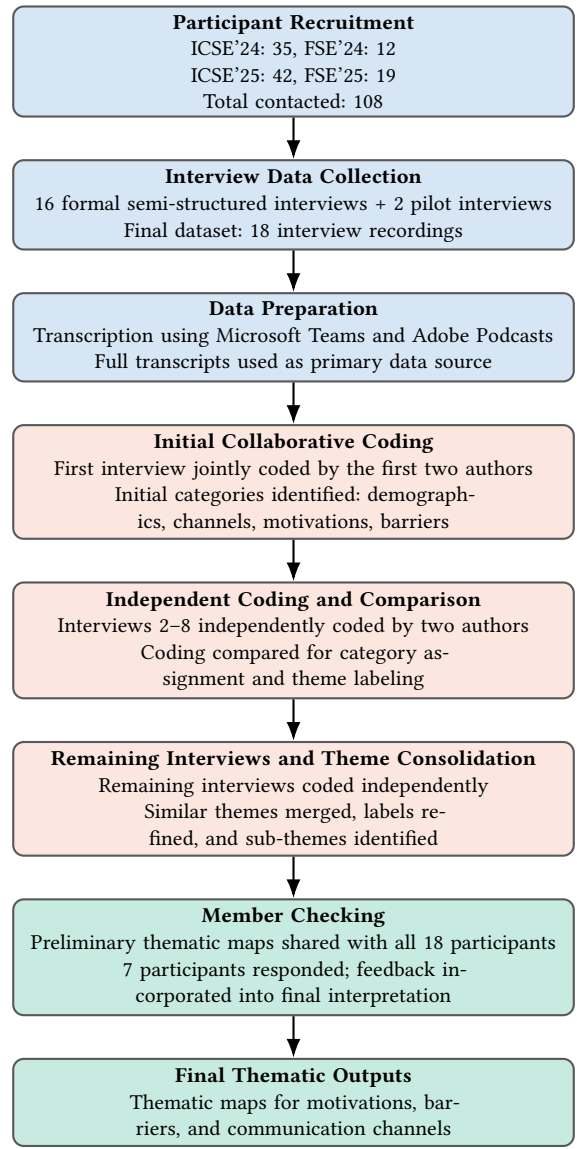

Each participant received a summary document synthesizing the collective insights derived from the interviews, rather than individual transcripts or codes. We asked participants to reflect on the summary and provide feedback by responding to three open-ended questions: (1) whether they recognized their personal experiences and impressions in the presented summaries, (2) which aspects they would emphasize based on their own experiences, and (3) whether any aspects of the findings were surprising to them. Participants were given a clear deadline for responding and were informed that their feedback would be used to refine and validate the study’s interpretations.
7 out of 18 participants responded to our member checking emails. This self-selected subset largely confirmed the thematic maps and indicated that the findings resonated with their experiences. The feedback did not lead to major structural changes in the analysis, but it helped us refine wording and clarify a few interpretations. While this process strengthened the credibility of the findings, the limited response rate also means that the feedback reflects only a subset of the original participants. Figure~\ref{fig:data-collection-analysis} shows the entire data collection-to-analysis process.

\subsection{Threats to Validity}
\label{sec:validity}

In qualitative research, several types of validity threats can influence the validity of findings. We discuss these in the context of our study and the steps taken to mitigate them.

\textbf{Construct Validity:} Construct validity concerns whether the study instruments and procedures accurately capture the phenomena under investigation. In this study, the phenomena are PhD students' motivations, barriers, and channels for science communication; the interview guide was grounded in prior work on science communication~\cite{burns2003science,besley2019strategic} and iteratively refined through pilot interviews to improve face validity and item clarity. However, two questions were added to the interview guide after the pilot phase, so the two pilot participants were not exposed to those items; this created a small asymmetry in topic coverage that may have reduced the observed prevalence of themes directly prompted by those questions (notably the Career Advancement motivation theme regarding academia-versus-industry distinctions and career benefits). To mitigate bias introduced by this asymmetry, all transcripts were coded collaboratively by multiple authors and themes were validated through coder discussion and consensus; nevertheless, readers should interpret prevalence estimates of the affected theme with this limitation in mind.

\textbf{Internal Validity:} 
Internal validity, often discussed as credibility in qualitative research, concerns whether the interpretations faithfully represent participants’ perspectives and experiences. To improve credibility, two authors independently coded the interview data and subsequently reconciled discrepancies through structured discussions. Themes were developed iteratively, enabling continuous refinement and validation of emerging interpretations. We further strengthened credibility through member checking, where preliminary findings were shared with participants for feedback. The responses supported the consistency between our interpretations and participants’ intended meanings.

\textbf{External Validity:} External validity concerns the generalizability of findings beyond the studied sample. Our participants were recruited from ICSE and FSE Doctoral Symposium tracks, representing a selected, research-active subgroup of PhD students whose work was accepted at major international venues and who chose to engage with the broader SE community.
As such, the sample may over-represent individuals who are already relatively confident and motivated in communicating their research. We acknowledge that our findings may not fully represent the experiences of all SE PhD students, particularly those from institutions or regions underrepresented in these conferences. Nevertheless, by including participants from diverse geographic regions and stages of the PhD process, we managed to capture a broad range of perspectives.

\textbf{Reliability:} Reliability in qualitative research refers to the consistency and repeatability of the study procedures and coding. To enhance reliability, we documented all steps of the data collection and analysis processes, including coding rules, theme definitions, and decisions made during thematic reconciliation. Multiple authors participated in coding and cross-checked each other’s work, reducing the likelihood of idiosyncratic interpretations. Furthermore, our coding and thematic frameworks are fully documented and made publicly available on Zenodo~\cite{chakraborty2026replication}.

By acknowledging these validity threats and taking concrete steps to mitigate them, we aim to provide a transparent and trustworthy account of how SE PhD students perceive and engage in science communication.

\section{Results}
\label{sec:results}

In this section, we present the results of our data analysis based on the coding of 18 interviews, addressing our three research questions (RQs).
We also included the two pilot interviews in the analysis, as they contained rich data and no substantial changes were made to the study design after the pilot phase. Thus, their inclusion did not introduce noise into the dataset.
%

\subsection{Participant Demographics}

\begin{figure}[t]
    \centering
    \includegraphics[width=\linewidth]{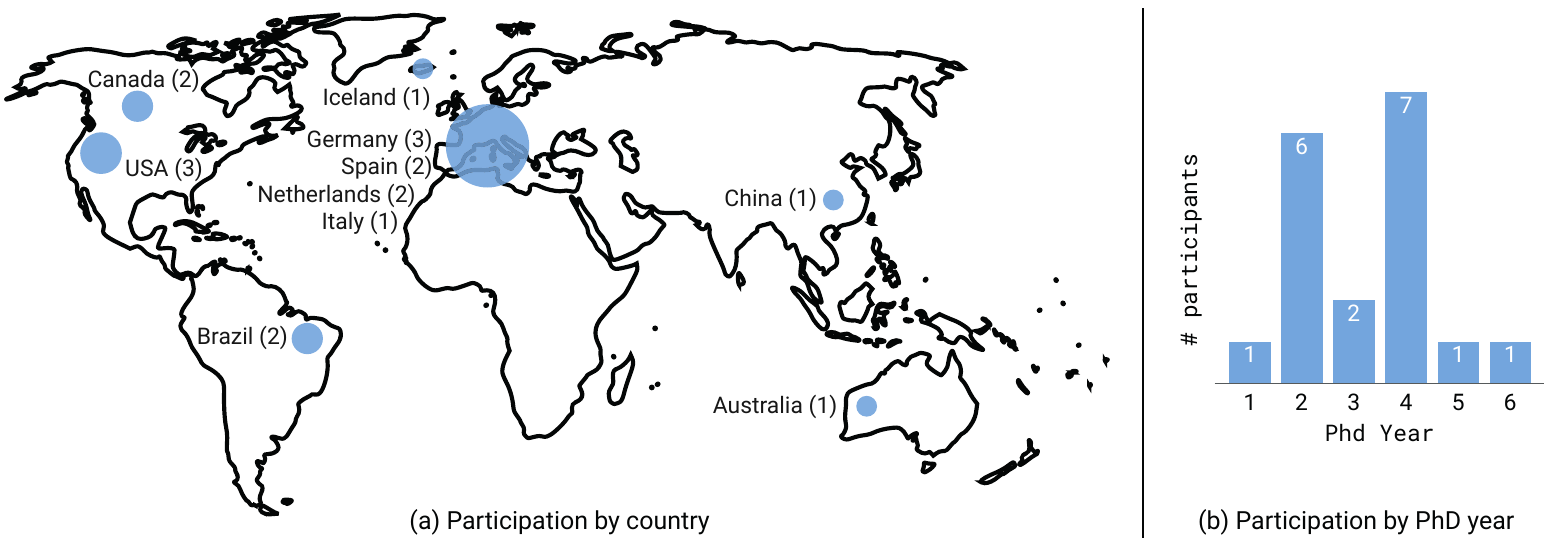}
    \caption{Overview of participant demographics.}
    \label{Fig:participant-demographics}
    \Description{Firstly, we see the geographical distribution of our participants across different continents, which corresponds to a fairly global coverage with a focus on North America and Europe. Secondly, we see the distribution by PhD year, with most of our participants being in their second or fourth PhD year at the time of the interviews.}
\end{figure}

We begin by presenting the demographic characteristics of our participants. 
Figure~\ref{Fig:participant-demographics} summarizes participation by university location and PhD year. 
Our sample includes students from diverse geographic regions, reflecting a global representation of software engineering doctoral researchers.
Some of the participants also left their home countries to pursue their PhD studies abroad, allowing them to provide additional insights into perceived differences in communication behavior across different cultures.
Participants were at different stages of their doctoral journey, ranging from first-year students to those close to completion, providing a balanced perspective on communication practices across (early) career stages.

\subsection{What motivates SE researchers to share their research?}

\begin{figure}
    \centering
    \includegraphics[width=1\linewidth]{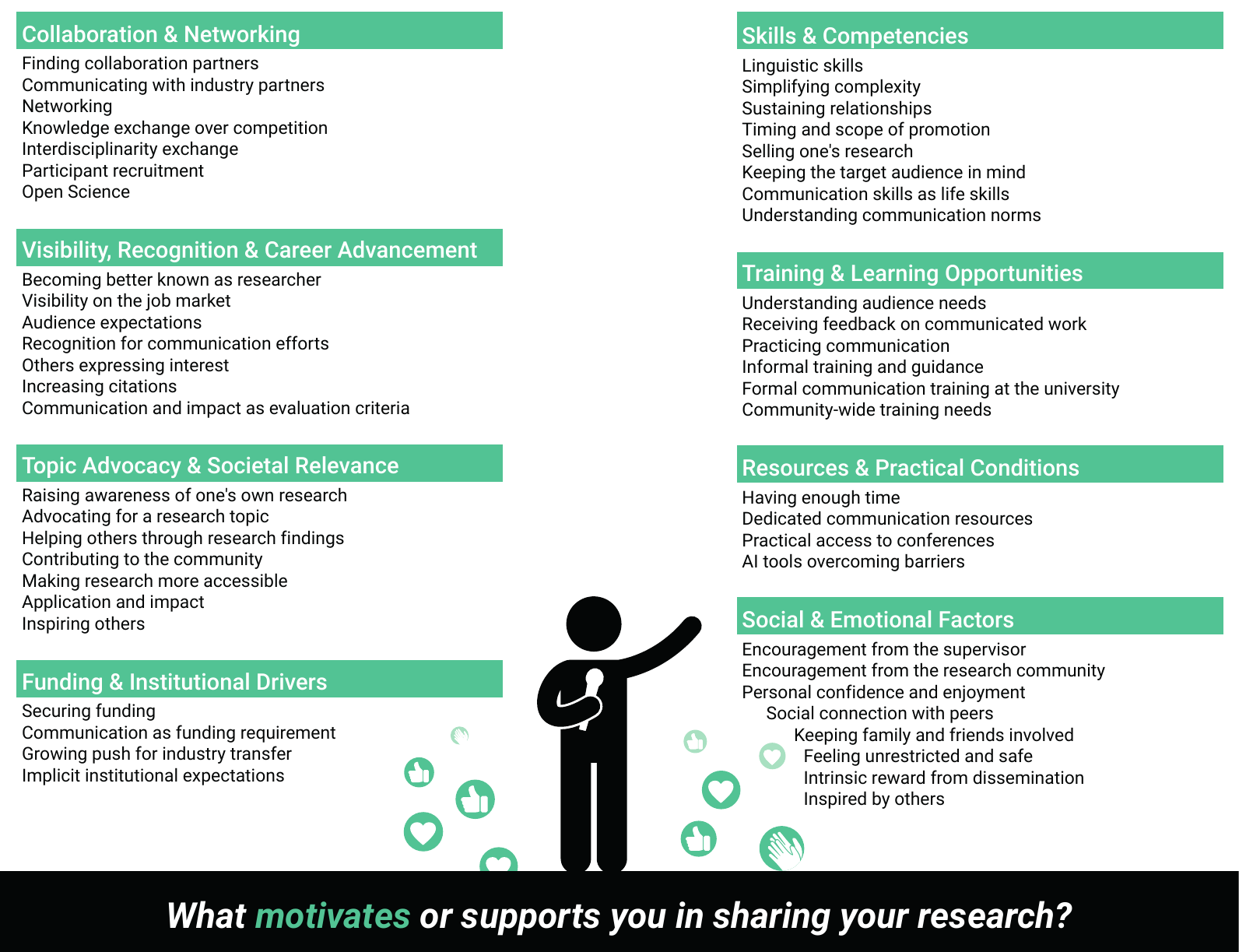}
    \caption{Thematic Map: What motivates or supports SE researchers in sharing their research?}
    \Description{Thematic map of motivations for sharing research, organized into eight themes: topic advocacy and societal relevance, collaboration and networking, visibility/recognition and career advancement, funding and institutional drivers, skills and competencies, training and learning opportunities, resources and practical conditions, and social and emotional factors. Each theme branches into its constituent codes.}
    \label{fig:map-motivations}
\end{figure}

From the perspective of our participants, there are numerous reasons that motivate and support deliberately communicating one's own research to one or more target audiences.
We summarize these reasons in Figure~\ref{fig:map-motivations} using eight themes comprising a total of 51 codes.
Across themes, these motivations were widely shared among participants.
Most themes, such as Collaboration and Networking, Social \& Emotional Factors, Training \& Learning Opportunities, Skills \& Competencies, as well as Visibility, Recognition \& Career Advancement, were supported by accounts from nearly all participants (16--17 out of 18), indicating a high level of shared experience.
Themes related to Topic Advocacy \& Societal Relevance (14 out of 18), Resources \& Practical Conditions (10 out of 18), and Funding \& Institutional Drivers (7 out of 18) were mentioned less often but still represent perspectives from a substantial part of the sample.
In the following, we discuss each theme and include excerpts from the interviews that are particularly characteristic of our participants' perspectives within the respective theme.

\subsubsection{Topic Advocacy \& Societal Relevance}

We start with a motivation that may be among the most intrinsic: participants genuinely believe in the importance of their work and that sharing it could benefit others.
Across seven codes, their accounts reveal a strong sense of conviction that communicating research is part of responsible scholarship.


\paragraph{Raising awareness, advocating, and inspiring others}
A first cluster of motivations reflects the participants' desire to make their work and topics visible, advocate for their importance, and spark interest in others. 
Several participants emphasized that research only matters if it becomes known.
As P1 noted, \enquote{if you are not communicating about your research you can make \ldots great information, but nobody's going to know.} 
Similarly, P4 reflected self-critically that they \enquote{really should [share more], because that's how research gets communicated.}
Beyond visibility, participants often expressed genuine passion for their topics.
P2 stated plainly, \enquote{I like to talk about my research because I'm passionate about it.}
P14 stressed that audiences \enquote{need to hear} about the subject, and P18 added, \enquote{it's really something everybody should know.}
This advocacy was closely tied to moments where communication visibly resonated with others.
For instance, P17 described seeing \enquote{the light bulbs going in their head,} highlighting how inspiring others can itself become a reinforcing motivation.

\paragraph{Helping others and giving back to the community}
A second group of motivations centers on prosocial intentions.
Participants frequently framed sharing as a way to directly help people or reciprocate support from their communities.
P1 described being \enquote{very motivated by helping people,} adding that it would feel unsatisfactory if only their immediate project team benefited from the PhD work.
Others emphasized community reciprocity: P7, for example, explained a desire to \enquote{give a feedback for them. Like thank you for helping me and this is what I found in your community.}
These accounts suggest that communication is often viewed not merely as dissemination, but as an ethical obligation to return value to those who enabled the research.



\paragraph{Making research more accessible and having an impact}
A third cluster highlights motivations related to translating research into accessible forms and enabling practical use.
Participants reported actively seeking ways to make their work more digestible for broader audiences.
P4 contrasted blog posts written \enquote{in layman's terms} with \enquote{stuffy academic writing,} while P18 emphasized wanting \enquote{to make [computer science] more accessible to the outside world.}
Closely connected was the aspiration to see research applied in practice.
Participants repeatedly linked communication with impact: \enquote{You want to have a sense of impact, right?} (P4).
Others stressed the importance of practitioners being able to use their work in day-to-day contexts (P7) and expressed hope that their research would be \enquote{actually applied in the industry}~(P5).
In this view, communication is a necessary step to ensure that research remains \enquote{impactful and practical \ldots instead of being just an article} (P9).



\noindent\\Taken together, the theme \emph{Topic Advocacy \& Societal Relevance} captures a deeply value-oriented motivation in which sharing research is seen as essential for visibility, societal benefit, and practical impact.

\subsubsection{Collaboration \& Networking}

Next, we explore a theme capturing participants' drive to build connections and exchange knowledge.
Across seven codes, participants emphasized that communication is about forming relationships, learning from others, and building networks that support ongoing research engagement.


\paragraph{Building collaborations and professional networks}
A first cluster of motivations centers on connecting with others to create collaborative opportunities. Participants often highlighted that sharing their work allows them to identify partners for joint projects or industry collaborations.
P13 noted that communicating ideas enables researchers to \enquote{bounce off more ideas in the context of what you do, what they are doing \ldots finding more collaborations.}
P9 similarly described reporting to industry as a way to identify potential collaborations: \enquote{maybe we can get some potential fundings or potential collaborations with them in some projects.}
Networking was also framed as essential for ongoing engagement with peers.
P4 observed that engaging with others \enquote{can \ldots lead to networking, getting connections, and future research projects,} while P17 stressed the importance of informal interactions, noting that \enquote{people throw names at you like all the time \ldots the community does value it quite a lot.}
Collectively, these accounts portray sharing research as a way to establish and strengthen professional connections.

\paragraph{Knowledge exchange and interdisciplinary dialogue}
Another important motivation was the opportunity to exchange knowledge and engage across disciplines. Participants emphasized that sharing research is a way to learn from others, reduce unnecessary competition, and stimulate innovation.
As P2 explained, \enquote{I want them to know what I'm doing, I want to know what they are doing \ldots see if they found a solution to that.}
P1 similarly highlighted the value of exchanging ideas even when collaboration does not occur: \enquote{it's good to know what people are doing. Also share like some resources.}
Interdisciplinary exchange further broadens this motivation: P5 noted that talking to people from other fields \enquote{improves the overall dynamics,} and P15 described adapting presentations to \enquote{gain more audience from different fields.}
Together, these insights show that communication fosters an environment of collaborative learning and cross-disciplinary enrichment.

\paragraph{Participant recruitment and Open Science}
Finally, participants emphasized motivations tied to practical requirements and institutional context.
Sharing research often serves concrete purposes, such as recruiting participants for studies or projects.
P3 described needing to \enquote{engage more with all the, you know, the stakeholder groups \ldots} to involve participants, and P16 noted the importance of industry contacts for evaluation purposes.
Institutional and structural factors also encourage open communication.
P5 explained that owning intellectual property and facing no university restrictions makes it easier to share work freely, while P18 highlighted the necessity of communication in contexts where data is scarce: \enquote{we have a lot of problems with data. We don't have any data. So for me, it's very, very important to communicate with others.}
These examples illustrate that contextual factors can strongly motivate researchers to engage in open and collaborative communication.



\noindent\\In sum, participants view sharing their research less as one-way dissemination and more as a means to connect, learn, and collaborate with others.

\subsubsection{Visibility, Recognition \& Career Advancement}
While the previous theme emphasized connection, exchange, and collaborative engagement, this next theme about \emph{Visibility, Recognition \& Career Advancement} highlights how communication can shape researchers' professional recognition and visibility within and beyond academia.

\paragraph{Establishing oneself in the research community}
A first cluster reflects participants' desire to become better known in their research communities and to develop an independent academic identity.
Some participants described the need to \enquote{put yourself in the front} because academic careers require personal scholarly presence and \enquote{people need to know you and not your advisor} (P1).
P11 pointed out that being \enquote{not a known researcher in the community} requires greater effort to get their work recognized by the community.
Career considerations were also visible in participants' accounts.
Several participants associated communication with future employment opportunities, particularly when approaching the job market.
For example, P4 and P12 mentioned that promoting papers becomes more important when planning to search for positions, while P9 described communication activities as a way to reach researchers who may recruit for postdoctoral roles.
\emph{Visibility} was a frequent keyword in this regard, e.g. \enquote{if you want to stay in academia, become professor, it's important to be visible} (P16).
In some cases, increasing citations was mentioned as one indicator of scholarly attention and academic influence.

\paragraph{Audience expectations and recognition for communication efforts}
A second cluster relates to how participants adapt their communication practices in response to audience behaviour and interest.
Some participants expressed uncertainty about whether traditional publications are widely read, which motivated them to use more accessible formats.
For instance, P14 noted that audiences may not read long papers and therefore used visual summaries such as posts on LinkedIn to convey main ideas.
Participants also valued situations where others showed curiosity about their work.
P16 described how it felt positive when people were interested in understanding their research and willing to listen to explanations of their PhD topic.
Recognition of communication efforts also appeared in interpersonal feedback contexts, such as encouragement received after presentations, suggesting that appreciation from peers can reinforce motivation.
In general, our participants felt confident that communication efforts are appreciated by the community: \enquote{And they recognize that, you know, [\ldots] what's the point of research if nobody reads it} (P4).

\paragraph{Communication and impact as evaluation criteria}
Finally, participants discussed how dissemination activities are increasingly embedded in formal evaluation systems.
Some noted that external communication and outreach can be considered in project or institutional assessments.
P10 mentioned that dissemination is now evaluated by some commissions, reflecting growing institutional attention to research impact beyond publications.
They also suggested that communicating research can indirectly support academic success by demonstrating practical relevance to practitioners and reviewers.
    

\noindent\\We see that participants associate research communication with professional recognition, career development, and academic influence. 
Sharing research thus serves to help researchers position themselves within scholarly and professional environments.

\subsubsection{Funding \& Institutional Drivers}

While visibility and recognition can function as personal rewards, participants also described how funding structures, institutional contexts, and broader strategic priorities encourage research communication.

\paragraph{Securing funding}
A first motivation relates to accessing financial resources and supporting research activities through communication.
Participants suggested that being active in communication can open pathways to collaboration and funding.
P9 described reporting research activities to industry as a way to \enquote{get some potential fundings or potential collaborations.}
Similarly, P11 highlighted that making research known can create more opportunities for projects, financial support, and resource acquisition.
This perspective was also reflected in institutional contexts of large applied research organizations with strong industry orientation, where maintaining contact with industry partners was described as important for supporting application-driven research and funding acquisition.

\paragraph{Institutional requirements and emerging strategic priorities}
Participants also noted that communication is sometimes motivated by explicit or implicit institutional expectations.
In some cases, outreach and dissemination are formal components of grant applications or project reporting.
For example, P17 described grant programs that required researchers to communicate research goals and progress as part of project proposals.
More broadly, participants observed a growing emphasis on industry transfer and practitioner-oriented dissemination in their research environments.
P8 noted that their institute aims to strengthen industry transfer, while P10 described dissemination as increasingly oriented toward practitioner audiences.

\noindent\\These perspectives suggest that \emph{Funding \& Institutional Drivers} represent a more structurally embedded motivation for communication.
Beyond individual goals, researchers also engage in sharing research to align with funding mechanisms, institutional priorities, and broader strategic trends toward knowledge transfer.

\subsubsection{Skills \& Competencies}

We now turn to a set of themes that focus less on external rewards or strategic opportunities and more on the personal capabilities that support research communication. The theme \emph{Skills \& Competencies} captures participants' reflections on the abilities and practices they consider important for communicating research effectively.
A central idea running through this theme is the notion of \enquote{selling one's research,} which many participants mentioned in different forms, often linked to presentation style, audience awareness, and strategic communication practice.
While the remaining codes were discussed less extensively, they nevertheless illustrate complementary aspects of communication competence.

\paragraph{Linguistic skills}
Participants highlighted linguistic skills as an enabler of research communication, particularly when working across international or multicultural research environments.
Strong language proficiency, such as English or bilingual abilities, was described as facilitating interaction with other researchers and communities.
P9 observed that individuals with stronger English skills or more extroverted personalities may be more inclined to communicate effectively.
P5 added a cultural touch, finding it easier to connect with other PhD students from the Asian region due to their own background.

\paragraph{Simplifying complexity}
Beyond language proficiency, participants emphasized the ability to simplify complexity.
Effective communication was associated with distilling research ideas into clear, high-level explanations.
For example, participants admired communication styles that used sketches or narrative summaries to make topics \enquote{very, very clear} (P1) and stressed the importance of \enquote{making complex things in a simple way} (P15).
Several participants criticized overly detailed presentations were \enquote{nobody gets it} (P18), suggesting that clarity and focus are valued communication characteristics.



\paragraph{Sustaining relationships and communicating strategically}

Another important aspect concerns sustaining professional relationships and understanding when and how to promote research.
Participants viewed the ability to keep past collaborations active as part of good scientific communication and community building.
P2 described examples of researchers who maintain long-term connections, while P11 highlighted individuals who effectively \enquote{sell} their ideas by communicating with different companies and partners.
It was not enough to have a good idea or solution; one also had to communicate its value effectively: \enquote{It got in the direction of actually pitching. Selling your research project for collaboration} (P8).
Strategic timing was also mentioned.
Participants observed that sharing new publications or research results soon after publication can be effective, with examples of researchers posting work on social media platforms such as LinkedIn: \enquote{He has a great timing. So as soon as he has published a paper, he is putting it on social media} (P1).
These practices reflect an understanding that communication is not only about content but also about presentation context.

\paragraph{Keeping the target audience in mind and understanding communication norms}
Participants further emphasized the importance of keeping the target audience in mind when communicating research.
Several noted that communication style should vary depending on the community or purpose of the message.
P2 suggested that authors should already consider their intended audience when writing papers, adjusting emphasis on certain points accordingly.
Similarly, P4 described tailoring messages when addressing audiences outside one's core research community by including actionable recommendations.
Awareness of communication norms within specific research communities was also mentioned.
Participants indicated that successful communication requires understanding how ideas are typically framed in presentations and discussions within their field.
Industry collaborations play a special role in this context, where certain approaches have proven successful: \enquote{So we are just putting what our industry partner said as problem up there and trying to go from there. And usually that's also how we communicate in presentations} (P8).
    


\paragraph{Communication skills as life skills}
Finally, participants described communication competence as a broader personal capability rather than a purely academic skill.
Some viewed communication skills as valuable beyond research contexts, describing them as general life skills.
P18 expressed that being able to communicate effectively is a meaningful long-term personal goal, while P1 associated communication practice with becoming more comfortable presenting oneself and one's ideas.


\noindent\\Taken together, the findings suggest that participants view research communication competence as a combination of technical language ability, simplification of complex ideas, strategic presentation, audience awareness, and relationship management.
These skills are perceived as integral to effectively \enquote{selling} research ideas and engaging with audiences of different backgrounds.

\subsubsection{Training \& Learning Opportunities}

We now turn to a theme that captures how communication is something researchers learn, refine, and actively develop over time.
Participants described communication as a space for gaining insight, receiving feedback, and building competence through both formal and informal learning experiences.

\paragraph{Understanding audience needs}
Several participants described communication as a way to better understand the needs and realities of their audiences.
Engaging with practitioners or other stakeholders helped them see differences between assumptions in research and what actually happens \enquote{outside of the academic bubble} (P6).
P1 noted that \enquote{there is some stuff that I see on industry that I don't see in research,} highlighting how dialogue reveals gaps between theory and practice. 
Others emphasized that understanding real-world problems is essential for meaningful research.
As P13 put it, \enquote{you can only solve problems for other people if you know what problem they're facing.}
In this sense, communication functions as a learning opportunity that sharpens researchers' awareness of whom they are actually addressing.


\paragraph{Receiving feedback on communicated work}
Participants also highlighted the importance of receiving feedback when they share their research.
While peer review offers one type of evaluation, several participants stressed the value of feedback from their target audience.
P2 explained that \enquote{to encourage the application of my tool in practice, I would also need some feedback from practical users.}
Others described critical feedback as constructive and motivating.
P3 noted being \enquote{happy to receive their criticism because I need to improve as well,} and P16 described feedback from industry confirming that \enquote{we're on the right track} as \enquote{really amazing.}
Science communication thus becomes a mechanism for validating and refining research ideas.


\paragraph{Practicing communication}
Opportunities to practice communication were mentioned as valuable for building confidence and skill.
Participants described public talks and presentations as rehearsal spaces.
P9, for instance, viewed public communication as preparation for a PhD defense, while P13 described iterative attempts to become \enquote{a better storyteller} when presenting research findings.
Several participants emphasized that repetition reduces anxiety and improves performance.
As P17 reflected, early conference presentations can be stressful, but over time the experience becomes more routine: \enquote{I think just the general experience of doing it regularly. I mean, I'm sure your first conference... you were sweating bullets and now like you next time you're finishing [your presentation] in the last minute just before you go on the stage.}    
Practice was therefore seen as central to developing effective communication habits.
    

\paragraph{Informal training and guidance}
Beyond individual practice, informal mentoring played an important role.
Supervisors were frequently mentioned as sources of advice and encouragement.
Participants described rehearsing presentations with supervisors, receiving guidance on how to write emails, or being provided with examples of effective dissemination materials.
P3 explicitly stated, \enquote{I learned quite a lot from him,} referring to supervisory support.
This informal guidance often reduced stress and provided a safe environment to experiment with communication formats.


\paragraph{Formal communication training and community-wide training needs}
Participants described structured training formats---such as university courses, workshops, and doctoral symposiums---as valuable opportunities to develop communication skills.
These settings addressed questions like identifying the audience, clarifying the core message, and making presentations engaging rather than tedious.
For example, P9 referred to a symposium in which senior researchers provided concrete guidance on \enquote{how to nail job talks and poster presentations}, which was perceived as highly helpful.
Similarly, courses on communicating research introduced different venues and formats, as well as strategies for tailoring content to specific audiences.

At the same time, several participants suggested that such training should be more widely integrated into academic practice.
They observed that many talks and reports fail to maintain audience attention and argued that researchers should systematically learn how to present their work \enquote{more effectively and more clearly} (P9).
This reflects a broader perception that communication training is not merely an optional add-on, but a skill set that would benefit the research community as a whole, supporting clearer dissemination, stronger engagement, and greater accessibility of research beyond specialist circles.
    


\noindent\\The theme \emph{Training \& Learning Opportunities} highlights that research communication is closely intertwined with processes of learning and skill development.
Through interaction, feedback, mentoring, and structured training, researchers gradually build the competence and confidence required to share their work effectively.

\subsubsection{Resources \& Practical Conditions}
Our next theme covers how structural conditions and the availability of resources shape researchers' ability to engage in communication activities.

\paragraph{Having enough time}
The availability and prioritization of time emerged as a central practical condition.
Some participants explicitly stated that having enough time enables them to prepare communication activities, such as presentations or outreach formats, even though it may happen during the later hours of the day: \enquote{I mean, I work my hours during the day and then I have enough time to prepare my communication presentations and all these kind of things} (P11).
Others framed time as a matter of strategic decision-making.
P16 compared communication to marketing, asking how much time should be spent \enquote{advertising} versus {building the product.}
Similarly, P17 emphasized that it depends on how communication is prioritized.
Rather than describing time as fundamentally lacking, participants often portrayed it as something that can be allocated intentionally within the broader scope of research work.


\paragraph{Dedicated communication resources}
Institutional support in the form of dedicated communication resources also played an enabling role.
Participants mentioned infrastructure such as institutional social media channels, support staff, or communication teams that help disseminate research outputs.
For example, tagging institutions on platforms like LinkedIn or Instagram could amplify visibility (P11), and some institutions actively supported researchers by preparing or reposting content: \enquote{I'm sure if you have an article and give it to the institute, they'll be very interested to put it out as their Facebook post or whatever} (P17).
In some cases, groups had even hired specialized staff for tasks such as graph visualization, which participants interpreted as a strong commitment to communication.
Such dedicated communication resources reduce individual effort and signal that dissemination is institutionally valued.


\paragraph{Practical access to conferences}
Access to conferences, including funding for travel and registration, was another important condition.
Several participants noted that their institutions covered travel expenses and provided financial resources to attend events.
This support made it at least easier, if not possible at all, to present work, connect with others, and engage in professional exchange.
The availability of funding thus directly enabled participation in communication-oriented academic settings.


\paragraph{AI tools overcoming barriers}
Finally, AI tools overcoming barriers were mentioned as a technological resource that lowers communication thresholds.
In particular, P15 referred to AI as helpful in addressing language-related challenges, making it easier to write posts or prepare written communication.
They noted, with current AI tools, \enquote{that to write a simple post is not a difficult part.} 
Such tools were seen as reducing practical barriers and facilitating more confident participation in communication activities (e.g. on social media).


\noindent\\Taken together, the theme \emph{Resources \& Practical Conditions} underscores the importance of enabling conditions in shaping communication practices.
Participants described time allocation, institutional infrastructure, financial support for conferences, and AI-based tools as concrete factors that either facilitate or constrain their engagement.
Rather than focusing on personal dispositions, these accounts point to the practical arrangements that make communication sustainable within everyday research routines.

\subsubsection{Social \& Emotional Factors}
We turn to a theme that is foregrounding the human dimension of research communication.

\paragraph{Encouragement from the supervisor}
Encouragement from supervisors emerged as one of the most prominent factors.
Participants described supervisors recommending conference participation (P6), stressing the importance of talking to senior researchers (P9), and actively encouraging the sharing of findings (P10).
Some supervisors prompted concrete actions, such as maintaining a personal website (P10), rehearsing presentations together (P17), or ensuring that PhD students present their work to improve communication skills: \enquote{If one PhD is with him, he will or she will present it because he wants us to improve our communication skills} (P11).


\paragraph{Encouragement from the research community}
Beyond supervisors, encouragement from the research community also played a motivating role.
Positive comments after presentations, supportive reviewers, and kind interactions during poster sessions were described as affirming experiences.
P5 recalled how a session chair expressed encouragement after a presentation.
P15 reported that both conference audiences and reviewers appreciated a simple and accessible writing style.
Even critical feedback was often perceived as constructive rather than hostile: \enquote{I feel, at least in academia, they are like very kind and sweet, while they're trying to criticize you} (P17).
Such reinforcement strengthened participants' willingness to continue sharing their work.


\paragraph{Personal confidence and enjoyment}
Personal confidence and enjoyment were frequently mentioned.
Many participants described themselves as comfortable with public speaking or even as particularly confident compared to peers.
For example, P1 thought they were \enquote{pretty okay} right now when it comes to presenting, especially since they consider themselves to be \enquote{a very talkative person, so I really enjoy talking about my interest.} 
Similarly, P2 stressed that they \enquote{never had much of a problem talking in public} and P4 consider themselve to be \enquote{definitely one of the more confident ones, if not the most confident} among their peers.
Several participants explicitly stated that they enjoy presenting or find it \enquote{good fun} (P12).
Others expressed satisfaction in showcasing their work and receiving feedback.
In these accounts, communication is not an obligation but rather an activity that aligns with personal disposition and provides positive emotional experiences.


\paragraph{Social connection with peers}
Social connection with peers was another important aspect.
Communication formats such as interactive sessions or informal discussions were described as opportunities for bonding and mutual understanding.
P13 pointed out that PhD life can be isolating, and such exchanges help counteract that isolation.
P17 emphasized that discussing research in slightly informal settings feels particularly fulfilling.
Science communication thus contributes to a sense of belonging within the academic community.


\paragraph{Keeping family and friends involved}
Keeping family and friends involved also motivated communication.
Participants mentioned explaining their work to parents (P3), updating friends from school (P4), or sharing research progress as part of their broader life experience (P12).
Even when explanations were challenging---such as when a parent \enquote{doesn't get it} (P8)---the effort reflected a desire for mutual understanding.
Communication here extends beyond professional audiences and connects research to personal relationships.


\paragraph{Feeling unrestricted and safe}
\enquote{I never had a feeling that [\ldots] something is stopping me from sharing my own research}~(P5). 
This sense of feeling unrestricted and safe was central for several participants.
They described environments in which they could communicate freely without fear of negative consequences: \enquote{If I want to communicate I communicate it} (P6).
The perception of having sufficient resources and skills to \enquote{put out stuff if we really want to} (P17) further contributed to this sense of autonomy.
This is where the \emph{Skills \& Compentencies} as well as the \emph{Resources \& Practical Conditions} themes appear to be enabling factors to this feeling of unrestrictedness.


\paragraph{Intrinsic reward from dissemination}
Intrinsic reward from dissemination was also evident.
For some participants, promoting their research felt less like additional work and more like a satisfying outcome of prior effort.
As P5 put it, compared to the research itself, dissemination was \enquote{not an effort. It's a reward.} Communicating results provided a sense of accomplishment and completion.


\paragraph{Inspired by others}
Finally, being inspired by others motivated communication activities.
Observing peers or senior researchers writing blog posts, delivering strong presentations, or adopting recognizable formats such as the \enquote{better poster} approach\footnote{https://osf.io/ef53g/files/6ua4k} encouraged participants to experiment with similar practices.
Role models within the community thus served as catalysts for PhD students' own science communication engagement.


\noindent\\We have already seen this shine through in a few of the other themes, for example when it came to skills or career aspects, but here it becomes clear once again that there is a person behind every communication effort.
And whether and how this person chooses to share their work depends on encouragement, confidence, belonging, safety, and inspiration.

\subsection{Which channels do SE researchers use to communicate their research?}
\label{sec:channels}

\begin{figure}
    \centering
    \includegraphics[width=1\linewidth]{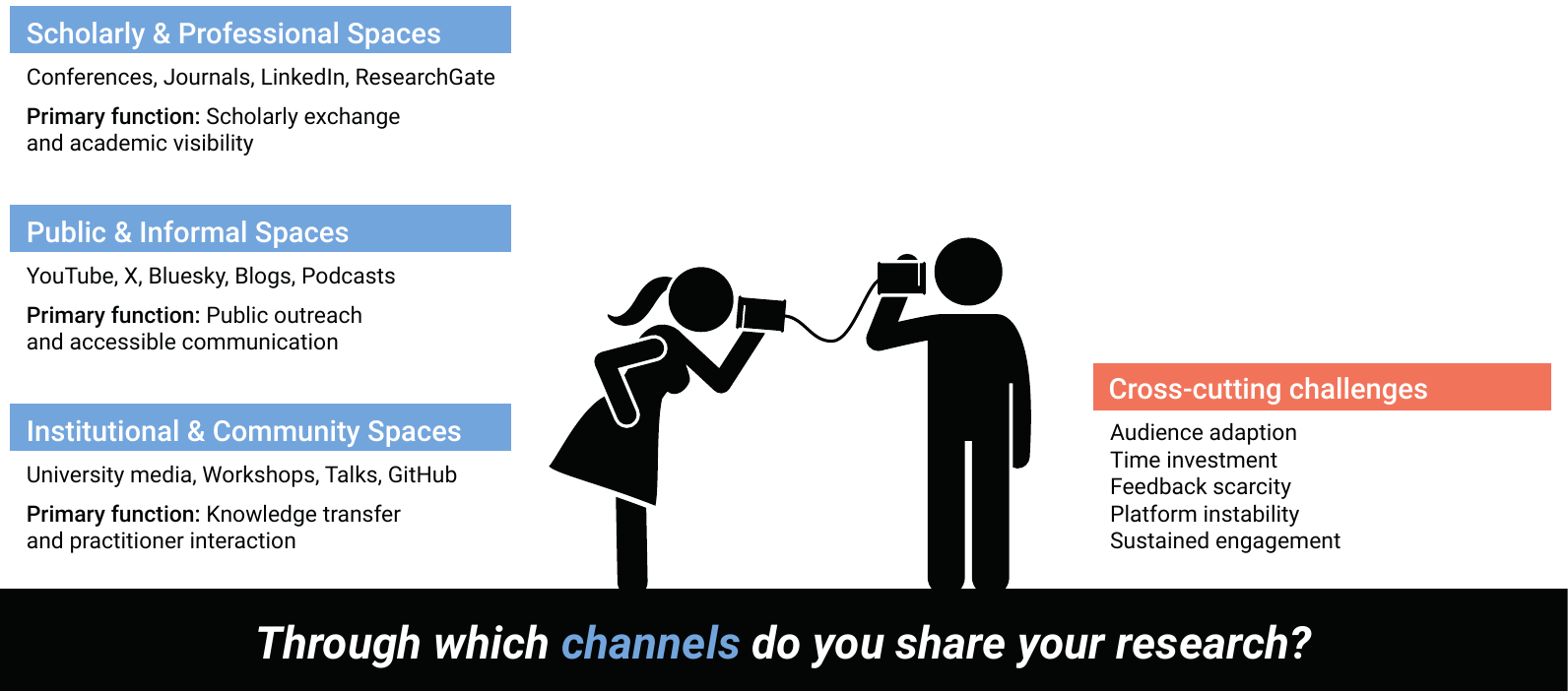}
    \caption{Communication channels identified in our study. Participants described complementary channel categories with distinct communication goals while sharing common practical challenges.}
    \Description{Overview of the communication channels participants reported, grouped into categories: academic and professional channels (conferences, journals, ResearchGate, arXiv, ORCID, LinkedIn), social media and informal channels (X, Bluesky, Facebook, Instagram, YouTube, blogs, podcasts), and institutional and community-based channels (university media, workshops, talks, GitHub, developer communities).}
    \label{fig:communication-channels}
\end{figure}

All 18 participants reported using, or having prior experience with, multiple communication channels throughout their research careers. In this study, we define a \enquote{communication channel} as any medium through which researchers disseminate or discuss their work. To answer this research question, we extracted all interview segments in which participants referred to communication channels and systematically categorized them according to the type of medium discussed (see Figure~\ref{fig:communication-channels}). While some channels were widely adopted across participants, others appeared only sporadically, indicating that researchers rarely rely on a single communication avenue. Instead, participants frequently combined multiple channels depending on their intended audience, communication goals, and available opportunities.

\subsubsection{Scholarly and Professional Spaces}
Participants frequently described scholarly communication channels as the foundation of research dissemination. Traditional scholarly venues, particularly conferences and journals, were consistently identified as central mechanisms for not just communicating research findings, but  participants viewed these channels as essential for presenting methodological contributions and establishing credibility. As P17 noted, \enquote{I don't write paper thinking of trying to communicate with people. It's just making one solid, rigorous piece of literature}. Similarly, P8 described publication as the core mode of scholarly presence in the community: \enquote{Way how to do it for me is is writing papers and submitting papers}. 

In contrast, professional communication channels were primarily associated with networking, broader reach, and more immediate visibility. Among these, LinkedIn emerged as the most commonly used platform.  P1 explained, \enquote{Right now I have my LinkedIn}, while P10 noted that they had moved away from X and now used LinkedIn for \enquote{the entire paper}. Several participants explicitly contrasted LinkedIn with more academic or less active platforms. P2 observed that many people in the academic community do not use LinkedIn much, but still considered it especially useful for industry partners, and P3 argued that \enquote{LinkedIn is more like the academic environment than other social media}, making it a more effective place for scholarly posts. Participants also described LinkedIn as a space for presenting research in more digestible forms; for example, P14 noted that posters on LinkedIn can help audiences understand \enquote{the main issues} without reading lengthy papers. These statements show that professional platforms were not seen as substitutes for scholarly publication, but rather as complementary channels that help research travel further and connect with practitioners and wider professional audiences.

Several participants additionally referred to research-oriented platforms such as ResearchGate, arXiv, and ORCID. Although these platforms were mentioned less frequently, they were commonly used to maintain an academic presence, improve accessibility of publications, and increase the visibility of research outputs beyond individual publication venues. As P16 explained, \enquote{you have your ORCID ID, which is I think a good thing. Theres everything centralized}, while P18 noted that \enquote{publishing stuff on archive really helps}.

Taken together, participants described scholarly and professional channels as serving complementary roles. While scholarly channels primarily support formal dissemination and scholarly credibility, professional platforms appear to extend the visibility of research and enable interactions with broader professional communities.

\subsubsection{Public and Informal Communication Channels}
A second theme captures the use of social media and other informal channels to communicate research in more accessible, interactive, and audience-oriented ways. Several participants reported using platforms such as YouTube, X, Facebook, Instagram, or Bluesky to share insights about their work based on accessibility. For example, P7 explained that these platforms are attractive because they are often free to use: \enquote{Like LinkedIn, your ResearchGate and YouTube or other communities, most of there are free to share. So I try to use them.} Similarly, P10 noted that \enquote{Sometimes Instagram} is used when presenting something at a conference, suggesting that these platforms are sometimes treated as extensions of formal communication rather than as fully separate spaces.
Participants also emphasized that public platforms often require concise and carefully framed messages. P17 described Bluesky as a platform where researchers need to \enquote{compress it to like few words and make it very clear and make it interesting.} This points to a communication style that differs markedly from academic writing, as the message must be brief, immediately understandable, and attention-grabbing. Likewise, P13 suggested that Twitter can support this kind of compressed communication, observing that it may be easier to communicate a paper if one has \enquote{the best tweet}. At the same time, P4 described Twitter threads as a way of turning a paper into a more accessible summary by \enquote{putting the main points} into a layman’s version of the paper.

Video and blogging formats were also mentioned as effective ways of explaining research concepts in a more narrative and visual way. P4 described blog posts as \enquote{a little bit more in layman's terms, you know, not the stuffy academic writing}, while P15 expressed a desire to write a blog that would explain research \enquote{in a simple way} and help people better understand what they were working on. In a similar vein, P13 noted that they had \enquote{always considered making short videos on YouTube}, although they ultimately felt that it involved \enquote{too much effort}. These accounts suggest that such formats are valued because they can simplify complex ideas and make research more accessible to non-specialist audiences.
However, adoption of these platforms varied considerably among participants. Some described limited engagement, while others noted a shift in platform preferences over time. P17 observed that \enquote{everyone has migrated, at least the big shots}, and P10 similarly explained that researchers are \enquote{moving from X, from Twitter to LinkedIn}. P2 also reflected on this transition, noting that Twitter had been used by the research community but that this had changed as \enquote{Elon Musk \dots has been kind of not being as active}. These comments indicate that public communication channels are shaped not only by affordances for sharing research, but also by changing platform cultures and community migration.

Taken together, these channels illustrate how Public spaces and informal media can complement more formal dissemination channels by enabling broader visibility, less formal interaction, and more accessible forms of communication. Rather than replacing scholarly venues, they appear to extend research communication into spaces where audiences can encounter, interpret, and discuss research in more immediate and flexible ways.

\subsubsection{Institutional and Community Spaces}
A third group of channels includes communication supported by institutions, communities, and organized events. Participants reported sharing their research through university communication resources, workshops, seminars, talks, and technical platforms. University-supported channels were often used to showcase research achievements through internal colloquia, institutional databases, or communication staff. For example, P1 noted that \enquote{in my internal university, we also have the colloquium so we can try on and share the research}, while P3 explained that supervisors often promote work because \enquote{they have more connections} and can help attract wider attention. P12 similarly described \enquote{smaller, more local events} within or near the university as useful opportunities to discuss research.
Workshops were described as especially interactive settings for exchanging ideas and presenting results to targeted audiences. P3 said that \enquote{if there's a chance we could conduct some kind of workshop, I can present some result. I think that would be much better}, and P10 suggested that \enquote{organized workshops or little events with practitioners} could be a good way to to reach industry partners than papers. Similarly, P11 noted that companies often create their own workshop venues at conferences, while P16 argued that once tools are integrated into practice, a workshop can support more direct exchange with vendors. P17 also described workshop-like seminar series as \enquote{formal, slightly informal interactive sessions} that were more fulfilling than one-way communication.

Talks and seminars were similarly framed as valuable oral formats for sharing research in smaller, more conversational settings. P1 referred to internal colloquia as a place to \enquote{share the research}, and P17 described seminar-style meetings as occasions where people \enquote{just talk among selves}. These settings were valued for enabling discussion rather than only formal presentation. P17 also noted that retreat-style events can create space for \enquote{informal chats} within a department.
Finally, technical and community platforms such as GitHub or GitLab were mentioned as ways to share software artifacts and related research outputs. Although these platforms were not described as central communication channels, P1 recalled using the GitHub education page for live streams, and P16 described the idea of a GitHub page as part of an unfinished personal website project. Together, these examples suggest that institutional and community-based spaces support communication by combining formal structures, collaborative interaction, and practical dissemination beyond traditional publication venues.

\subsubsection{Communicating Research Effectively Across Channels}

Beyond the choice of channels, participants emphasized that effective communication also depends on how research is presented. 
Several participants highlighted the importance of narrative structure, particularly in presentations. 
As P4 noted, effective presentations often resemble storytelling, where a clear narrative guides the audience and connects theoretical contributions to real-world contexts.
Participants also emphasized the importance of adapting communication strategies to different audiences. 
For example, P1 explained that presentations designed for researchers may differ substantially from those aimed at practitioners, noting that \enquote{you really need to adjust. It's not just copy paste the presentation.} %
Written communication was similarly discussed as an important aspect of science communication. 
Participants stressed that accessible and clearly written papers play an important role in making research understandable to a broad audience. 
P4 highlighted that a well-written paper, particularly one that is accessible for readers such as non-native English speakers, can itself constitute an effective form of communication.
Participants further noted that simplicity and visual elements can help make research more approachable. 
For example, P15 described using interactive presentations to explain complex concepts in simpler terms, while P1 referred to examples where research ideas were summarized visually through sketches or drawings.

Overall, these reflections suggest that participants view science communication not merely as selecting appropriate channels, but as a broader practice involving narrative framing, clarity of presentation, and adaptation to diverse audiences.

\subsubsection{Practical Considerations and Limitations of Communication Channels}
While participants described many opportunities for sharing their research, they also reflected on practical limitations associated with different channels. 
One recurring theme concerned the balance between formal academic dissemination and everyday communication practices. 
Although conferences were widely recognized as effective venues for structured communication, some participants perceived them as less relevant for day-to-day engagement. 
As P12 reflected, conferences \enquote{do a good job in terms of facilitating that kind of communication, but on a day-to-day basis, I don't know that it's a priority.}
Maintaining profiles across multiple platforms was also described as effort-intensive. 
For example, P16 mentioned that although they had created a profile on ResearchGate, they rarely update it due to time constraints.
Cost considerations also influenced channel preferences. 
Several participants noted that they primarily rely on free platforms to share their research. 
As P7 explained, platforms such as LinkedIn, ResearchGate, or YouTube are attractive because \enquote{most of these are free to share.}
Finally, participants reflected on potential demographic differences in platform usage. 
For instance, P17 observed that certain communication spaces, such as blogs within their field, appeared to be more male-dominated.

Taken together, these observations highlight that the choice of communication channels is shaped not only by their potential reach but also by practical constraints, evolving platform dynamics, and the effort required to maintain an active presence.

\subsection{What barriers do SE researchers face when communicating SE research?}

From the perspective of our participants, there are numerous barriers that make communicating their research challenging. 
These barriers range from difficulties in identifying and reaching appropriate audiences to personal, social, and structural constraints that hinder effective science communication. 
Social barriers were reported by the majority. Ten out of the 18 participants mentioned language, cultural differences, and social anxiety as major challenges. Eleven participants described a lack of guidance and training opportunities for communication, while seven reported a lack of institutional resources to support them. Another common challenge was understanding and communicating complex software engineering topics in a way that resonated with others. Without proper guidance, as mentioned by most participants, we observed genuine motivation that was often constrained by these barriers.

\begin{figure}
    \centering
    \includegraphics[width=1\linewidth]{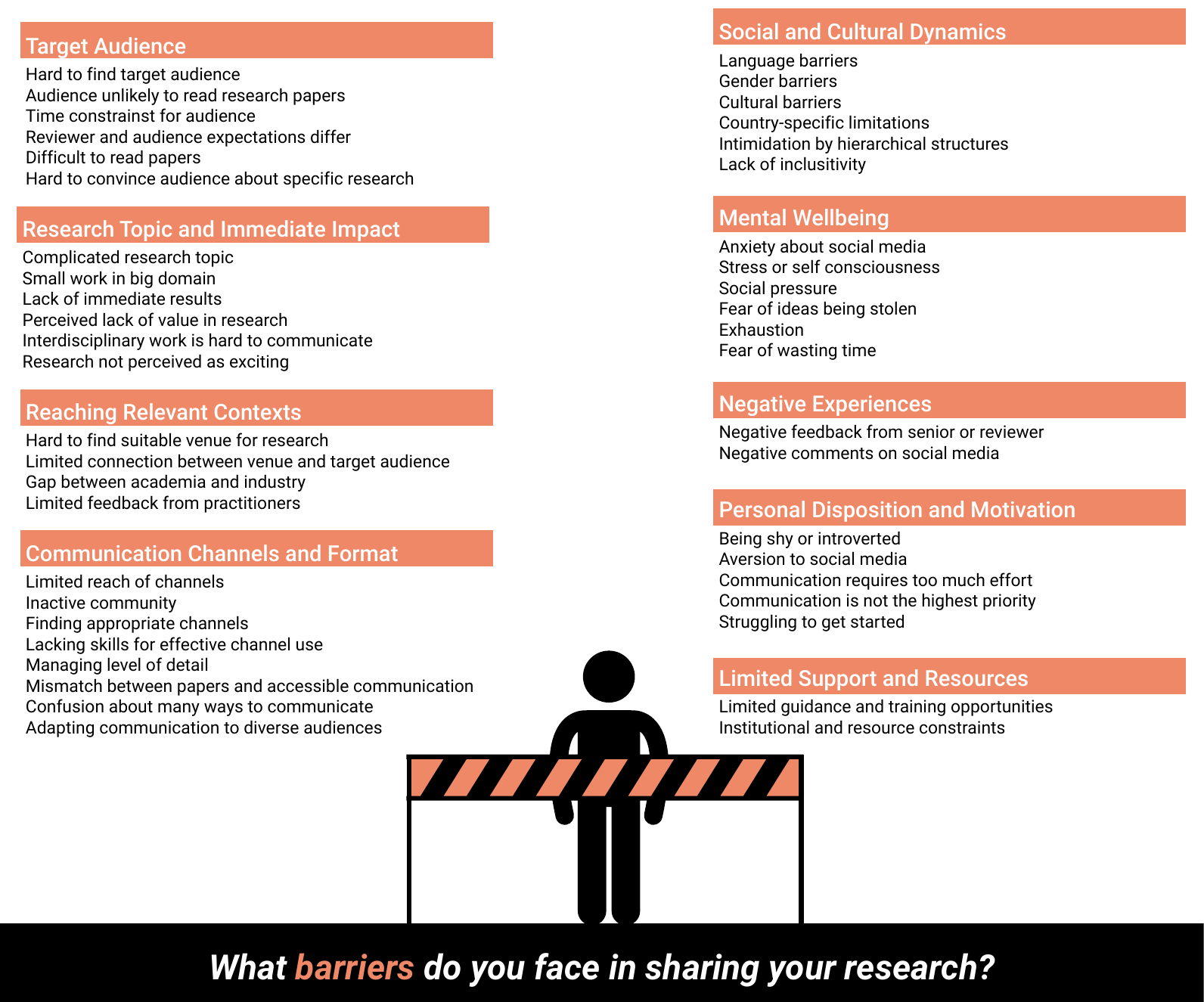}
    \caption{Thematic Map: What barriers do SE researchers face in sharing their research?}
    \Description{Thematic map of barriers to sharing research, organized into ten themes: target audience, research topic and immediate impact, reaching relevant contexts, communication channels and format, social and cultural dynamics, mental wellbeing, negative experiences, personal disposition and motivation, and limited support and resources. Each theme branches into its constituent codes.}
    \label{fig:map-barriers}
\end{figure}

We summarize these barriers in Figure~\ref{fig:map-barriers} using nine themes comprising a total 45 codes. 
In the following, we discuss each theme in turn and present excerpts from the interviews that are particularly characteristic of our participants' perspectives within the respective theme.

\subsubsection{Target Audience}

A first major barrier relates to identifying and effectively communicating with the intended audience for research. 
Participants frequently described uncertainty about who their work is actually reaching and whether the people who could benefit from it are able or willing to engage with academic research outputs.

\paragraph{Hard to find target audience}

Several participants reported difficulties in identifying the specific individuals or communities that might benefit from their research. 
Even when research targets industry or practitioner communities, researchers may struggle to locate the relevant stakeholders within organizations or professional networks. 
As P8 reflected, \enquote{then you understand that it's actually just two or three people in the company that actually interested in what you're doing and you just hope that it works out.}
This uncertainty makes it difficult to tailor communication strategies and ensure that research reaches those who might find it useful.

\paragraph{Audience unlikely to read research papers}

Another recurring barrier concerns the format of academic publications themselves. 
Participants noted that many of their intended audiences—particularly practitioners—do not regularly read research papers or attend academic conferences. 
For example, P2 explained that a large part of their intended audience consists of industry professionals \enquote{who may not really be reading my papers or going to this kind of conference,} particularly in domains such as automotive or aerospace engineering. 
Similarly, P18 suggested that many practitioners would not engage with traditional research papers because \enquote{it needs to be on a higher abstraction level.}

\paragraph{Time constraints for audience}

Participants also emphasized that practitioners often lack the time required to engage with lengthy academic publications. 
Even when research might be relevant, the time investment required to read and understand a full paper can discourage potential readers. 
As P3 observed, \enquote{I don't think they have time to read my paper, even if they have time, they wouldn't read such long papers.}
These constraints further limit the reach of traditional academic communication formats.

\paragraph{Reviewer and audience expectations differ}

Several participants described tensions between the expectations of academic reviewers and those of their intended audiences. 
While reviewers often expect detailed methodological explanations and formalized presentations of research problems, practitioners may prefer concise and actionable insights. 
P4 described attempting to address this challenge by connecting their research to practical recommendations in the discussion section, explaining that they try to \enquote{speak directly to my target audience if possible, and have some actionable items for them as well.}
However, as P18 noted, the increasing formalization required in academic writing can sometimes make research less engaging for broader audiences, remarking that \enquote{there's a lot of formalizing the problem, and I think at that point nobody listens anymore.}

\paragraph{Difficult to read papers}

Participants also pointed to the complexity of research papers as a barrier to effective communication. 
Technical language, dense explanations, and highly specialized terminology can make it difficult for audiences outside the immediate research community to understand the work. 
P2 noted that reviewers or readers may not always grasp the complexity of the challenges being addressed, stating that \enquote{sometimes the viewer do not understand the complexity of the challenges I'm addressing.}
P3 prefers application made after the results than paper, \enquote{If the day comes and then maybe I'll just present my results or even, you know, present in an application way like in an app}. Complex papers are barriers to understand certain topics and researchers might prefer talking to an expert or known peer rather than reading such papers, as mentioned by P17, \enquote{ I find a lot more information in 1/2 an hour talk I'll have with some person in some uni who works in a similar field. Than, like through any paper}. Such misunderstandings can limit the accessibility and perceived relevance of research outputs.

\paragraph{Hard to convince audience about specific research}

Finally, participants described challenges in persuading audiences of the relevance or importance of their research. 
Communicating research effectively often requires understanding the background knowledge, interests, and context of the audience—something that is not always easy to determine. 
P8 highlighted this difficulty, explaining that it can be \enquote{very hard to identify directly what their context is and what their background knowledge is and how much they can correctly identify what I'm talking about.}
Similarly, P1 emphasized that adapting the message itself can be challenging, noting that \enquote{adapting the message is the hardest part and making it more appealing.} 

\noindent
Taken together, the theme \emph{Target Audience} highlights how challenges related to identifying, reaching, and engaging the intended audience can significantly hinder the effective communication of software engineering research.

\subsubsection{Research Topic and Immediate Impact}

A second theme relates to barriers arising from the nature of the research itself and how its value or relevance is perceived. 
Participants explained that certain characteristics of research topics such as their technical complexity, narrow scope, or lack of immediate outcomes, can make them difficult to communicate effectively to broader audiences.

\paragraph{Complicated research topic}

Several participants described how the technical complexity of software engineering research can make it difficult to communicate to non-specialist audiences. 
Highly technical topics often require substantial background knowledge, which many audiences may not possess. 
As a result, researchers must simplify or abstract their explanations, which can itself be challenging. 
P10 noted that their work involves \enquote{a lot of technical parts} and explained that \enquote{if you are not an expert, it's difficult to communicate because in a lot of cases we work mainly on the back-end side.}
Similarly, P11 described attempting to present the main ideas at a higher level of abstraction but found that audiences without an engineering background still struggle to understand the concepts, explaining that \enquote{it's really complex for them to understand it because they are not engineers.}

\paragraph{Small work in big domain}

Another barrier arises from the perceived scale of individual research contributions within a broader domain. 
Participants noted that doctoral research often focuses on highly specific problems, which can make it difficult to convey the broader significance of the work. 
As P8 described, during a PhD \enquote{you're doing such a little tiny wheel in a big system,} or as P18 described having results too deep that it is hard to communicate their effects, \enquote{I know that I'm not addressing the people I want to because yeah, I think the results we're doing are going down too deep}, making it challenging to communicate why the contribution matters within the larger research landscape.

\paragraph{Lack of immediate results}

Participants also emphasized that research outcomes often become meaningful only over longer time horizons. 
This long-term nature of academic research can make it difficult to communicate immediate relevance or tangible benefits. 
For instance, P3 explained that some audiences expect short-term outcomes, whereas \enquote{a lot of research is focusing on the long-term perspective or long-term results.} 
Similarly, P6 suggested that this challenge is not unique to a single project but reflects a broader characteristic of academic work, noting that the difficulty in demonstrating immediate application is \enquote{a general problem of academia.}

\paragraph{Perceived lack of value in research}

Some participants also reported that certain types of research may be perceived as less valuable or impactful by their intended audiences. P6 questioned, \enquote{What is the impact? So you need to see the impact, how you can influence the society.}
When researchers feel that their work does not directly address the interests of practitioners or other stakeholders, they may instead communicate primarily within the academic community. 
For example, P12 said, \enquote{Importance and public interest are two different things}. Similarly, P18 reflected that their work may be \enquote{going down too deep,} which leads them to communicate mainly with other researchers. 
Similarly, P16 noted that their research topic \enquote{doesn't really change necessarily this much,} suggesting that incremental contributions may be harder to present as impactful to broader audiences.

\paragraph{Interdisciplinary work is hard to communicate}

Participants working across disciplinary boundaries described additional communication challenges. 
Interdisciplinary research often requires audiences to understand concepts from multiple domains, which can make explanations more complex and less accessible. 
For instance, P12 explained that their work intersects with legal research and therefore requires \enquote{intersectional knowledge to some extent,} making it more difficult to communicate to audiences who may not share expertise in both areas. 
Similarly, P14 noted that human-centered research topics can also be challenging to communicate within technical communities, explaining that they are working on \enquote{a human study topic} that differs from more traditional technical research areas.

\paragraph{Research not perceived as exciting}

Finally, several participants noted that some research topics may struggle to attract attention because they are perceived as less exciting or visually engaging than other areas of technology. 
For example, P8 contrasted their work on development tools with more attention-grabbing technologies, explaining that \enquote{development tool is not so shiny. It's not a general robotic device that brings you a robotic butler that brings you a cool drink, so it doesn't really get a lot of attention.}
On the same note, P11 feels, \enquote{maybe there is not much people interested on it (SE) than comparing for instance Distance astronomy or nuclear energy or things like that}.

\noindent
Taken together, the theme \emph{Research Topic and Immediate Impact} highlights how the inherent characteristics of research topics—such as complexity, scope, interdisciplinarity, and perceived novelty—can shape how easily they can be communicated to different audiences.

\subsubsection{Reaching Relevant Contexts}

Another barrier concerns the contexts in which research can be communicated and the environments where it may reach relevant audiences. 
Participants described difficulties in identifying appropriate venues, connecting with the intended communities, and bridging the gap between academic research and real-world practice. 
These challenges affect not only where research is shared but also whether it ultimately reaches those who could benefit from it.

\paragraph{Hard to find suitable venue for research}

Several participants reported difficulties in identifying suitable venues where their work could be communicated effectively. 
This challenge was particularly evident for research topics situated at the intersection of multiple fields, where established publication or communication venues may not clearly align with the work. 
As P2 explained, \enquote{it's hard to find a publisher there because you have only very few people that work on cyber physical systems, but it's also the problem of my field, which is a bit in the intersection between different areas.}
Participants also noted that communication venues often impose structural or procedural overhead that can make it harder to present results clearly. 
For example, P12 suggested that while venues aim to facilitate dissemination, it remains important to \enquote{allow people to get their results across clearly without too much overhead.}

\paragraph{Limited connection between venue and target audience}

Even when appropriate venues exist, participants observed that these venues do not always connect effectively with the intended audience. 
In particular, general software engineering conferences may attract broad research communities but may not include practitioners or specialists in specific subdomains. 
As P18 reflected, \enquote{the conferences we are addressing just isn't the right way to get a big community, because people usually on these conferences, it's more general software engineering and therefore they don't know about this concrete field.}
This misalignment between venue and audience can limit the reach and impact of research dissemination.

\paragraph{Gap between academia and industry}

Participants also described a broader structural gap between academic research and industrial practice. 
While researchers may aim to address practical challenges, practitioners often face different priorities, constraints, or timelines. 
As P6 explained, \enquote{we do our best, but in industry they have other problems to solve.}
Another gap is writing style. While academic writing style may not be enough to grab practitioners attention, researcher lack skill to adapt to industry style of writing, as mentioned by P10 \enquote{At research, in fact is very academic way of writing, because I think that if I acquire this skill, I can try to to convince practitioners to read my work.}
This divergence between academic interests and industry needs can make it difficult for research to resonate with practitioners or influence real-world practices.

\paragraph{Limited feedback from practitioners}

Finally, participants highlighted the lack of feedback from practitioners as a barrier to effective communication. 
Even when researchers attempt to share their work with practitioner communities, responses are often limited or absent. 
For example, P10 noted that despite sharing research results, they \enquote{never received feedback from practitioners, like an answer under the post or a comment,} with only one instance of feedback occurring during an in-person presentation. 
This absence of interaction can make it difficult for researchers to assess whether their work is relevant or useful to practice. 
Reflecting on this challenge, P6 expressed skepticism about the practical uptake of their work, stating that although they would like their research \enquote{to be applied in practice, realistically I understand it will never happen.}

\noindent
Taken together, the theme \emph{Reaching Relevant Contexts} highlights how structural and contextual barriers—such as difficulties identifying appropriate venues, misalignment between communication platforms and audiences, and limited engagement from practitioners—can hinder the effective dissemination and practical impact of software engineering research.

\subsubsection{Communication Channels and Format}

In Section~\ref{sec:channels}, we discussed the various channels participants use to share their research findings, as well as some of the limitations associated with these platforms. 
In this theme, we focus more specifically on the barriers that arise when using these channels, as well as challenges related to how research is formatted and presented. 
While communication channels determine where research is shared, communication formats influence how effectively the message is conveyed. 
Participants emphasized that both aspects are closely intertwined and jointly shape the success of science communication efforts.

\paragraph{Limited reach of channels}

A recurring concern among participants was the limited reach of individual communication channels. 
Even widely used platforms may not effectively reach all intended audiences, leading researchers to question the impact of their communication efforts. 
As P1 explained, \enquote{since X kind of died when I got the role, I started posting on LinkedIn to have more engagement,} while noting that alternative platforms such as Bluesky or Mastodon are not yet widely adopted. 
This highlights the difficulty of identifying channels that provide both visibility and broad audience coverage.

\paragraph{Inactive community}

Closely related to reach, participants also described challenges related to inactive or fragmented communities across platforms. 
Some channels may be suitable for specific audiences but lack engagement from others, particularly within the academic community. 
For example, P2 noted that although LinkedIn can be useful for industry partners, \enquote{there are a ton of people in the academic community who don't really use it much.} 
This fragmentation leads to uncertainty about whether investing effort in a particular platform is worthwhile, especially when no single channel effectively reaches all relevant audiences.

\paragraph{Finding appropriate channels}

Participants further highlighted the lack of a dedicated, widely adopted platform specifically designed for research communication. 
As P10 observed, \enquote{we don't have a platform like LinkedIn or Twitter specifically designed for researchers.} 
P10 also described LinkedIn as \enquote{a platform specifically designed for practitioners. So you know we there is this stranger behavior. And so somehow also this discourage me.} On the same note, P12 described it as \enquote{job hunting platform than a a social media platform.}
This absence of a clear, centralized communication space increases the complexity of selecting appropriate channels and aligning them with communication goals.

\paragraph{Lacking skills for effective channel use}

Beyond identifying suitable platforms, participants also described challenges related to effectively using communication channels. 
Creating engaging content for platforms such as YouTube, Instagram, or other media-oriented channels requires specific skills that go beyond traditional academic training. 
As P11 noted, while such formats may be effective for communicating research, \enquote{it's really hard work} to produce content that meets the expectations of these platforms. 
This highlights a skills gap between traditional academic communication and modern digital dissemination practices.

\paragraph{Managing level of detail}

In addition to channel-related challenges, participants described difficulties in determining the appropriate level of detail when communicating research. 
Simplifying complex ideas is necessary for broader audiences, yet oversimplification may risk losing essential nuances or reducing perceived rigor. 
As P18 expressed, \enquote{if I make it too easy, people will not see the abstraction, but rather think it's just easy stuff.} 
This tension between clarity and accuracy represents a central challenge in science communication.

\paragraph{Mismatch between papers and accessible communication}

Participants also pointed to a mismatch between traditional academic writing and more accessible forms of communication. 
While research papers often require formal and detailed presentations, broader audiences may benefit from simpler explanations and more engaging narratives. 
However, adapting content to be more accessible may conflict with academic expectations. 
For instance, P15 noted that reviewers may not always appreciate simplified explanations, even though they attempt to \enquote{explain in a simple way what the whole work is about.} 
This creates a tension between academic rigor and accessibility.

\paragraph{Confusion about many ways to communicate}

The wide range of available communication options can also create uncertainty about which strategies are most effective. 
Participants described feeling unsure about whether investing effort in certain channels or formats would meaningfully improve the reach of their research. 
As P16 reflected, \enquote{I'm not sure if making more LinkedIn posts is the way to do, because it's just reaching your bubble anyways.} 
This uncertainty can lead to hesitation or reduced engagement in communication activities.

\paragraph{Adapting communication to diverse audiences}

Finally, participants emphasized the challenge of tailoring communication to diverse audiences with different levels of expertise, interests, and expectations. 
Effectively communicating research often requires significant adaptation in language, framing, and format, depending on whether the audience consists of researchers, practitioners, or the general public. As told by P15, \enquote{And also you have the reviewers that don't like your way to writing}.
Balancing these differing needs adds complexity to the communication process and requires careful consideration of both content and delivery.

\noindent
Taken together, the theme \emph{Communication Channels and Format} highlights how challenges related to platform selection, audience reach, required skills, and the adaptation of content collectively shape the effectiveness of science communication efforts in software engineering.

\subsubsection{Social and Cultural Dynamics}

Beyond structural and research-related barriers, participants also described social and cultural dynamics that influence how researchers communicate their work. 
These challenges arise from linguistic differences, cultural norms, gender dynamics, and institutional hierarchies that shape participation within research communities. 
Participants emphasized that such factors can affect researchers' confidence, visibility, and ability to engage with broader audiences.

\paragraph{Language barriers}

Language proficiency emerged as a recurring challenge in science communication, particularly for researchers who are not native English speakers. 9 put of our 18 participants described language as a barrier for their experienced or perceived science communication. 
Since English dominates academic publishing and conferences, participants noted that limited language proficiency can make it more difficult to express ideas clearly and engage in discussions. 
For example, P4 acknowledged the advantage of being a native speaker, stating that they \enquote{benefit from being a native English speaker.}
In contrast, P9 explained that limited English proficiency can affect how effectively researchers communicate their ideas, noting that language skills influence their ability \enquote{to clarify their ideas clearly and to engage more discussions.}
At the same time, the effort required to communicate in a second language can be mentally exhausting. 
Reflecting on this experience, P9 described how they must constantly think about phrasing and expressions, explaining that \enquote{it's really nervous and really challenging because I always need to formulate all those kinds of expressions, and it's kind of exhausting.}
Similarly, P16 observed that researchers who struggle with English sometimes rely on highly scripted presentations, explaining that they may \enquote{prepare a presentation rather like a transcript and just read it out.}
These challenges can limit spontaneity, interaction, and confidence when communicating research.

\paragraph{Gender barriers}

Participants also pointed to gender-related dynamics that can influence interactions within research communities. 
For example, P18 noted that social interactions at conferences or networking events may sometimes occur along gender lines, explaining that \enquote{women usually approach each other, but it's rather difficult to get in touch with people which have not the same gender.}
Such dynamics can make it harder for some researchers to initiate conversations, build connections, or participate fully in professional discussions.

\paragraph{Cultural barriers}

Cultural norms regarding hierarchy and respect can also shape how researchers communicate with others in the academic community. 10 out of 18 participants talked about various spectrum of cultural barriers. 
In some contexts, strong respect for senior researchers may discourage junior researchers from actively engaging in discussions or presenting their ideas confidently. 
For instance, P9 explained that in their cultural context, \enquote{we have a deep respect to the senior researchers} which can make early-career researchers feel intimidated when introducing themselves or discussing their work. 
Similarly, P10 described hierarchical cultural structures in their country, explaining that \enquote{there is the boss and the boss is the absolute truth,} which can discourage open dialogue or critical discussion.
Being away from home, in a different culture makes one more private and restrictive to communication, as P2 said, \enquote{It can be sometimes a bit of a private thing. So even if you're facing challenges you usually don't broadcast, you may not want to broadcast it.}
The hidden challenge for majority is when communication feels like a burden of representation. P1 described it as \enquote{it can really feel like overwhelmed with the amount of invitation. And if you say no, you feel pressured because you're not representing your community}

\paragraph{Country-specific limitations}

Participants also described practical barriers associated with geographic and political constraints. 
For example, travel restrictions or visa issues can prevent researchers from attending conferences where they could present their work and connect with the community. 
As P17 explained, they \enquote{couldn't go to one conference because Canada didn't give a visa.} Another barrier is the current political situation as P4 described, \enquote{There's some privacy concerns, especially in this political climate.}
Such barriers can limit opportunities for international visibility and participation.

\paragraph{Intimidation by hierarchical structures}

Closely related to cultural norms, hierarchical structures within research communities can make it difficult for early-career researchers to gain visibility or speak confidently. 
Participants described feeling that senior voices dominate academic discussions, leaving limited space for junior researchers to contribute. 
As P11 explained, \enquote{right now it's hard to make sound in the community if you are young.}
P10 even mentioned the fear of \enquote{destroy my work in one minute maybe} talking freely to seniors with great expertise. 
This perception can discourage researchers from actively engaging in communication activities.

\paragraph{Lack of inclusivity}

Finally, participants highlighted broader inclusivity challenges within academic environments and conferences. 
These challenges may relate to accessibility, dietary restrictions, caregiving responsibilities, or other factors that affect participation. 
For example, P16 described how conference logistics can sometimes overlook diverse needs, noting that issues such as food restrictions, childcare responsibilities, or the presence of young children can make participation significantly more stressful. 
They reflected that these circumstances illustrate how \enquote{there are challenges where things are not equal.}

\subsubsection{Mental Wellbeing}

A particularly sensitive and critical barrier identified by participants relates to mental wellbeing. 
Across the interviews, participants consistently expressed concerns about the emotional and psychological impact of communicating their research. 
Unlike many other barriers, these challenges are deeply personal and often intertwined with confidence, self-perception, and social interaction. 
Participants frequently reflected on how anxiety, lack of confidence, and limited social skills influence not only their everyday academic experiences but also their willingness and ability to communicate their work effectively. 
These accounts suggest that science communication is not only a technical or strategic activity but also an emotionally demanding practice that can significantly affect researchers’ engagement.

\paragraph{Anxiety about social media}

Participants described experiencing anxiety related to engaging with social media platforms for research communication. 
This anxiety often stems from uncertainty about outcomes, perceived expectations, and comparisons with others. 
For example, P8 noted that they would feel anxious if they invested effort without seeing results, explaining that \enquote{I would feel this anxiety if I start now and I don't get anything out in the next 10 months.} 
Similarly, P5 reflected on feeling discouraged when comparing their own contributions to others, stating that \enquote{I feel lazy to draft something, and I felt like everyone was posting things that are so fancy.} 
Fear of being judged is another reflection from our participants, as P18 said, \enquote{this person posts weird stuff on LinkedIn and you really don't want to be part of that}
Such perceptions can discourage participation and reduce confidence in engaging with online platforms.

\paragraph{Stress or self-consciousness}
 
Participants highlighted that presenting ideas publicly involves a degree of vulnerability, particularly when engaging with audiences beyond their immediate research community. 
As P12 explained, \enquote{you're really making yourself vulnerable there,} emphasizing the emotional exposure associated with communication. 
In addition, concerns about audience reactions may lead to hesitation, as P14 noted a reluctance to engage more actively due to the fear that others \enquote{don't care,} which could further reduce motivation.

\paragraph{Social pressure}

Participants also reported experiencing social pressure related to expectations of maintaining visibility and actively communicating their work. 
This pressure can originate from both the academic community and broader norms around online presence. 
For instance, P9 described how the pressure to make a good impression can lead to nervousness, even when researchers are confident in their work. 
Similarly, P4 referred to a \enquote{constant source of anxiety} stemming from the feeling that they should be more active in sharing their work. 
These expectations can transform communication from a voluntary activity into a perceived obligation.

\paragraph{Fear of ideas being stolen}

Another concern relates to the perceived risk of sharing research ideas too early. 
Participants expressed caution about openly discussing ongoing work, particularly in informal or public settings, due to fears that others might appropriate their ideas. 
As P5 explained, there are cases where researchers \enquote{accidentally shared your novel ideas and then your audience maybe is not a friend of yours, and they scoop you.} 
This fear can limit openness and reduce willingness to engage in early-stage communication.

\paragraph{Exhaustion}

Participants also highlighted the physical and mental exhaustion associated with communication activities. 
Attending conferences, preparing presentations, and maintaining an active communication presence require significant energy in addition to core research tasks. 
For example, P5 noted that activities such as conferences can be \enquote{exhausting,} sometimes even leading to health-related consequences. 
Such experiences can discourage sustained engagement in communication efforts.

\paragraph{Fear of wasting time}

Finally, participants expressed concerns about the potential inefficiency of communication efforts. 
Investing time in activities that do not yield visible outcomes can lead to frustration and anxiety. 
As P5 reflected, interruptions or delays in research progress related to communication activities can create a sense of lost productivity, leading to concerns about \enquote{wasting time.} 
This perception can further reduce motivation to engage in communication.

\noindent
Taken together, the theme \emph{Mental Wellbeing} highlights how emotional, psychological, and social pressures can significantly influence researchers' willingness and ability to communicate their work, suggesting that effective science communication requires not only technical skills but also supportive environments that address these underlying concerns.

\subsubsection{Negative Experiences}

Few factors discourage researchers from communicating their work more strongly than negative experiences encountered in the past. 
Participants repeatedly reflected on how critical, dismissive, or even hostile interactions can shape their willingness to share their research. 
These experiences not only affect confidence but also influence how safe or worthwhile communication activities are perceived to be. 
Across the interviews, participants described negative encounters in academic settings, on social media, and in broader interactions, highlighting the lasting impact such experiences can have on their engagement with science communication.

\paragraph{Negative feedback from senior or reviewer}

A first barrier relates to negative feedback received from senior researchers, reviewers, or peers. 
While critical feedback is a fundamental part of academic practice, participants described situations where discussions became overly harsh, personal, or discouraging. 
For example, P10 highlighted the competitive nature of some research communities, noting that researchers may hesitate to speak up because they are \enquote{afraid that they (seniors) can answer you in a bad way} or get remarks like \enquote{it doesn't seem to me that these are constructive discussion. Your work does not make sense at all}.
In some cases, feedback was perceived as undermining the legitimacy of the research itself. 
P12 described an interaction in which their work was questioned at a fundamental level, being asked \enquote{are you sure this is a computer science PhD?} 
Such experiences can reduce confidence and discourage researchers from actively engaging in discussions or sharing their ideas.

\paragraph{Negative comments on social media}

Participants also pointed to negative interactions on social media as a barrier to communication. 
Unlike formal academic feedback, social media responses can be less structured and more emotionally charged, increasing the risk of confrontational or dismissive comments. As P9 mentioned, \enquote{I expect some discussions peacefully, but sometimes I don't know why the discussion, turns out to be a kind of attack and even kind of insult.}
The possibility of receiving public criticism or being misunderstood can make researchers hesitant to share their work in open online environments.


\noindent
Taken together, the theme \emph{Negative Experiences} highlights how past encounters with criticism, hostility, or dismissive attitudes can have a lasting impact on researchers' confidence and willingness to communicate their work.

\subsubsection{Personal Disposition and Motivation}

In addition to structural and contextual barriers, participants also described personal factors that influence their engagement in science communication. 
This theme reflects how individual dispositions, preferences, and priorities shape researchers’ willingness and ability to share their work. 
Although these barriers are more personal in nature, participants emphasized that they can have a substantial impact on communication practices and should not be overlooked.

\paragraph{Being shy or introverted}

Some participants highlighted that personality traits such as shyness or introversion can make communication activities more challenging. 
Engaging in discussions, networking, or presenting research—especially in the presence of senior researchers—may feel intimidating for individuals who are less comfortable in social or public settings. 
As P1 explained, when working closely with an advisor or in unfamiliar environments, being shy can make it \enquote{very difficult,} and may lead to situations where \enquote{your communication skills go down.} 
This suggests that confidence and comfort in social interactions play an important role in effective science communication.

\paragraph{Aversion to social media}

Participants also described a general reluctance or lack of interest in using social media platforms for research communication. 
For some researchers, these platforms do not align with their preferred communication style or professional identity. 
For instance, P12 noted that they \enquote{don't have a very large social media presence,} indicating limited engagement with these channels. 
Such preferences can reduce participation in widely used dissemination platforms.

\paragraph{Communication requires too much effort}

Another barrier relates to the perceived effort required for effective communication. 
Participants described communication as an additional task that requires time, preparation, and adaptation of content for different audiences, as P4 accepts, \enquote{I could be putting more effort into sharing, and I really should, because that's how research gets communicated.}
This includes simplifying complex ideas, maintaining online presence, and engaging with audiences across multiple platforms. 
For some researchers, these demands make communication feel burdensome, as P1 noted, \enquote{To be fair, like, I always wanted to write blog posts, but I think the effort is very big.} compared to other academic responsibilities.

\paragraph{Communication is not the highest priority}

Participants also emphasized that communication activities often compete with other priorities, such as \enquote{P1:  I need to take my time also to to do my research}, writing papers, or meeting project deadlines. Especially for a yong researcher or PhD students, as described by P2, \enquote{As a PhD student, you always have a ton of stuff to do.}
As a result, communication may be deprioritized, especially when it is not directly linked to formal academic evaluation or immediate career outcomes.

\paragraph{Struggling to get started}

Finally, some participants described difficulties in initiating communication activities. 
Even when motivated, researchers may delay or postpone engagement due to uncertainty, lack of time, or competing responsibilities. 
As P1 reflected, intentions to communicate can remain unrealized over time, noting that \enquote{I'm gonna do it, I'm gonna do it, and then the time passes,} eventually leading to missed opportunities to reach the intended audience. 
This highlights the challenge of translating intention into consistent communication practice.

\noindent
Taken together, the theme \emph{Personal Disposition and Motivation} highlights how individual traits, preferences, and competing priorities can influence researchers’ engagement in science communication, reinforcing that effective dissemination is shaped not only by external factors but also by personal circumstances.

\subsubsection{Limited Support and Resources}

Participants frequently described barriers related to insufficient institutional support and limited resources for science communication. While communication activities were often perceived as important or implicitly expected, researchers felt that the supporting structures necessary to engage effectively were frequently absent.

\paragraph{Limited guidance and training opportunities}

Several participants noted that they had received little guidance from supervisors, senior researchers, or institutions on how to approach science communication. In particular, early-career researchers described uncertainty about where to begin, which channels to use, and how to adapt communication efforts to different audiences. As P18 put it, \enquote{I don't know what's okay, what's not}. Without such guidance, communication was often experienced as a process of trial and error rather than a skill developed through structured support. P7 captured this bluntly: \enquote{No, you have to learn by yourself. Good luck.}

Participants also emphasized the lack of formal training opportunities dedicated to science communication. While academic training commonly covers writing papers or delivering conference presentations, communicating research to broader audiences often requires additional skills that are rarely addressed in doctoral education. P4 argued that \enquote{there should be some kind of class for PhD students} on how to communicate results, use social media, promote research, and apply for workshops and grants. P2 similarly pointed to the absence of structured preparation, saying simply, \enquote{So formal training. No,} while P1 highlighted a practical challenge: \enquote{the top one challenge that I have is making engaging}. They added that they were \enquote{very bad at doing graphics and images or summarizing what I've done}, underscoring how difficult it can be to translate research into accessible formats.

One participant also questioned the usefulness of available support, suggesting that \enquote{people are talking always about this coaching stuff, but it depends a little, but I think not everything is really helpful} (P18). Taken together, these accounts suggest that many researchers feel left to develop communication skills informally, with limited institutional support and few opportunities for structured learning.
\paragraph{Institutional and resource constraints}

Beyond guidance and training, participants also perceived institutional support for communication activities as limited. Some suggested that research communication was not viewed as a priority within their environments and that institutional incentives primarily focused on more traditional academic outputs. For example, P17 noted: \enquote{I don't think the university in general cares about at least our university because they're more oriented towards getting students and graduating them for industry. So I don't think they care a lot about communicating research.}

Practical resource limitations further constrained communication opportunities. Financial barriers such as conference travel costs and participation expenses were frequently mentioned. As P6 explained, \enquote{Everything cost money, to present something, cost money to travel somewhere, cost money. Someone needs to pay for it. That's a challenge.} Similarly, P2 noted that limited resources may restrict participation opportunities: \enquote{Of course, if I got accepted and had something to present because of course, like with this kind of international conferences, they cannot really afford to send everyone.}

Participants additionally emphasized time constraints as a persistent challenge. Communication activities often compete with research, teaching, and administrative responsibilities, leading communication efforts to be deprioritized despite their perceived importance.

Taken together, these findings suggest that communication barriers are not solely individual challenges but are also shaped by broader structural and institutional factors that influence researchers' ability to engage in science communication effectively.

\section{Discussion}
\label{sec:discussions}
Science communication for early-career software engineering researchers extends beyond the simple dissemination of research outcomes. Rather than functioning as a final stage after the research is completed, communication is embedded throughout the research process itself and intertwined with professional identity, collaboration practices, and career development. Across motivations, communication channels, and barriers, participants described communication as simultaneously desirable and challenging, revealing an ongoing tension between aspiration and practical engagement.

\subsection{Science Communication as a Socio-Technical Practice}
A recurring pattern in our findings is that participants rarely perceived science communication as a one-way transfer of knowledge. Instead, communication emerged as a broader socio-technical activity situated within interactions among researchers, practitioners, institutions, and communities. Participants frequently associated communication with collaboration opportunities, networking, interdisciplinary exchange, participant recruitment, and relationship building, suggesting that communication serves goals beyond just increasing visibility.
This observation aligns with the broader shift in science communication research away from traditional deficit-oriented models toward more dialogic and participatory approaches, where communication is viewed as an ongoing exchange rather than a simple dissemination of information~\cite{trench2008towards,besley2016qualitative}. Rather than appearing as a final stage after the research is completed, communication is perceived as embedded throughout the research lifecycle.
%

Within SE, this perspective may be particularly important because research frequently spans heterogeneous stakeholder groups including academics, practitioners, industry partners, and open-source communities. Existing work has emphasized the challenges of transferring knowledge between software engineering research and practice~\cite{koziolek2026dear,mikkonen2018continuous}.
%
Our findings suggest that science communication may indeed function as an important mechanism for bridging these communities and sustaining knowledge exchange.

\subsection{The Visibility--Vulnerability Tension}
Although participants strongly recognized the value of science communication, our findings reveal a substantial gap between acknowledging its importance and engaging in it confidently. Participants associated communication with recognition, visibility, impact, and career development, while simultaneously describing barriers such as social anxiety, fear of negative reactions, uncertainty regarding communication practices, and concerns about audience expectations.
These findings suggest a broader \emph{visibility--vulnerability} tension. Increased visibility creates opportunities for recognition, collaboration, and professional growth; however, becoming visible simultaneously increases exposure to criticism, judgment, and emotional pressure. Prior work on public engagement similarly reports that researchers often perceive communication activities to involve reputational and emotional risks, particularly when communicating outside traditional academic environments~\cite{poliakoff2007factors}. Communication therefore involve an ongoing balancing process in which researchers evaluate potential professional benefits against possible social and psychological costs.
%
This tension seems particularly pronounced for early-career researchers who are simultaneously developing professional identities and establishing positions within their research communities. Early-career researchers frequently experience uncertainty regarding expectations, self-presentation, and belonging within academia~\cite{gardner2008fitting}. 
%
Our findings extend these observations by suggesting that barriers to science communication are not simply a matter of lacking skills or motivation. Instead, science communication emerges as a complex practice that many doctoral students actively strive to engage in, as reflected by the eight motivation themes identified in our study. However, factors such as the complexity of the research topic, the surrounding environment, the availability of multiple communication channels, and students' social and mental well-being collectively create an uneven landscape that shapes their ability to communicate effectively.

\subsection{The Role and Complexity of Communication Channels}
One of the most prominent findings concerns the increasing diversity of communication channels available to researchers.
Our results provide a structured view by grouping channels into three distinct spaces: scholarly and professional, public and informal, and institutional and community contexts.
Participants reported using a broad ecosystem spanning traditional academic venues such as conferences and journals, professional networks such as LinkedIn, and various social platforms. Although this diversity creates new opportunities for dissemination and visibility, it simultaneously introduces additional complexity.
Prior research on communication environments suggests that channels are not neutral mechanisms for transmitting information but instead possess distinct affordances that shape interaction patterns, visibility, audience expectations, and communication practices~\cite{treem2013social,sitessocial}. Our own participants frequently described uncertainty about which channels are appropriate for particular audiences, how research should be adapted to different formats, and how much effort should be invested in maintaining multiple communication spaces.
%

In addition, participants reported challenges associated with maintaining engagement and obtaining meaningful feedback across platforms. Existing studies have similarly noted that digital communication environments can increase reach while simultaneously fragmenting audiences and creating uncertainty regarding impact and interaction~\cite{eysenbach2009infodemiology,DBLP:conf/icse/WyrichB24}.
By organizing this fragmented landscape into a small set of recurring channel types, our findings make these challenges more tractable, highlighting that difficulties often arise not from individual platforms but from navigating across spaces with different norms and audiences.
Consequently, the increasing number of communication opportunities may not necessarily reduce communication barriers. Instead, our findings suggest that researchers are increasingly faced with challenges related to channel selection, audience alignment, and the management of communication efforts across heterogeneous platforms.

\subsection{The Invisible Labor of Science Communication}
Participants repeatedly highlighted practical barriers including time constraints, limited institutional guidance, lack of structured training, and uncertainty about effective communication strategies. Collectively, these findings suggest that science communication often functions as a form of \emph{invisible labor}: work that requires substantial effort, yet frequently remains insufficiently recognized within formal academic structures.
Invisible forms of labor within academia have previously been associated with additional emotional and organizational burdens, particularly for early-career researchers who are simultaneously navigating professional expectations and developing academic identities~\cite{social2017burden}. 
%
Universities increasingly encourage researchers to engage with broader audiences and demonstrate societal impact, while evaluation systems continue to prioritize more traditional scholarly outputs such as publications, citations, and grants~\cite{madsen2019scientific}. Our findings reflect a similar tension. Producing communication materials, adapting content to different audiences, maintaining online profiles, and sustaining engagement across multiple platforms require considerable time and effort that compete with other responsibilities such as research, teaching, and administrative work.

\subsection{Communication as a Tool for Academic Visibility and Career Development}

Our findings suggest that science communication functions as a strategic resource for early-career researchers navigating a competitive academic environment.
Rather than framing communication purely as outreach, participants positioned communication as a means of establishing scholarly presence, signaling independence from senior collaborators, and becoming recognizable within relevant research communities.
This aligns with prior work on academic visibility and reputation-building, which highlights the importance of developing an identifiable research profile early in one's career~\cite{Majhi:2023:Practices}.
At the same time, communication practices were closely related to anticipated career transitions, such as entering the postdoctoral job market, where being known by peers and potential collaborators can shape opportunities: \enquote{if you want to stay in academia, become professor, it's important to be visible} (P16).

Importantly, these individual career-oriented motivations coexist with the view of communication as an entry point into intellectual exchange.
Participants also described participation as a way to access others' work and identify shared problems: \enquote{I want them to know what I'm doing. I want to know what they are doing \ldots see if they found a solution to that} (P2).
This reflects a more relational understanding of communication, consistent with perspectives that emphasize networked scholarship and the co-production of knowledge~\cite{Nicholas:2020:Attitudes,Park:2018:KnowledgeSharing}.
Interdisciplinary interactions further reinforce this dynamic, as adapting communication to different audiences enables researchers to traverse disciplinary boundaries and integrate diverse perspectives.
Taken together, these findings indicate that communication is embedded in both career-building and knowledge-building processes, functioning as a bridge between individual advancement and collective scientific exchange.

\subsection{The Need for Guidance and Institutional Support}
Our work represents an initial step toward understanding science communication within software engineering. Given the exploratory nature of our study and the relatively small sample of PhD students, our findings should not be interpreted as universally representative of science communication practices across disciplines or researcher populations. Rather than deriving general rules for science communication, our goal was to establish an empirical understanding of how early-career software engineering researchers experience communication in practice. This foundational understanding is necessary before effective interventions, support mechanisms, and communication strategies can be designed.

Importantly, the characteristics of software engineering itself shape the motivations and barriers identified in our study. Software engineering differs from many disciplines in several ways. Conferences play a particularly central role as publication and networking venues, strong interactions with practitioners and industry are common or at least desired, and communication frequently spans heterogeneous stakeholders, including researchers, software developers, educators, open-source communities and industrial partners~\cite{DBLP:conf/sigsoft/IvanovRSYZ17}. Furthermore, SE is a relatively young and rapidly evolving discipline that continuously adapts to emerging technologies and work practices. These characteristics may create unique communications opportunities while simultaneously increasing complexity in terms of audience selection, communication channels, and expectations about impact.
%

Recent work in science communication increasingly argues against the assumption of universally effective communication approaches and instead emphasizes context-dependent perspectives on what works for whom and under which circumstances~\cite{Achiam2025Terroir,besley2019strategic}. Similarly, software engineering research has repeatedly highlighted the importance of context when designing processes, methods, and interventions~\cite{dybaa2012works}.
Our findings reinforce this perspective and suggest that science communication may similarly require context-sensitive approaches rather than universal strategies.
%

With this work, we took a first step by providing an empirical foundation for understanding science communication as a context-dependent and socio-technical practice within software engineering. Consequently, future work should move beyond identifying barriers and begin investigating a broader objective: \emph{understanding which communication practices are effective, for whom, and under which contexts}. Potential directions include:

\begin{itemize}
    \item \textbf{Structured mentorship:} integrating guidance from supervisors and experienced researchers on when, where, and how to communicate research.
    
    \item \textbf{Communication training:} extending doctoral education beyond academic writing and presentations toward broader communication skills such as audience adaptation, storytelling, and online engagement.
    
    \item \textbf{Recognition mechanisms:} acknowledging communication activities as meaningful scholarly contributions rather than supplementary work.
    
    \item \textbf{Community-specific strategies:} investigating which communication practices are effective for different audiences, disciplines, and communication goals.

   \item \textbf{Mental well-being and psychological safety:} creating inclusive and psychologically supportive environments in which researchers can communicate without fear of excessive judgment, negative experiences, or social pressure. Such environments may help reduce the visibility--vulnerability tension identified in our findings and encourage more confident and sustainable engagement in science communication.
\end{itemize}


\section{Conclusion}
\label{sec:conclusion}
Strengthening science communication among PhD students can play an important role in making software engineering research more visible, accessible, and relevant beyond its immediate academic context.
Through an in-depth qualitative study with 18 software engineering PhD students across diverse contexts, our work provides insights into their lived practice of science communication.
This was a necessary first step in examining the factors that motivate or discourage researchers from talking about their research---an endeavor that is widely agreed to be important for the impact of the software engineering research field.
We highlight that fostering meaningful science communication requires not only skill development on the side of the PhD students, but also supportive environments that address knowledge of communication, enable feedback, and lower barriers to participation.
With strong motivation already in place, fostering the right conditions can help PhD students more fully realize their potential as communicators of their research.

\newpage
\bibliographystyle{ACM-Reference-Format}
\bibliography{refs}

\end{document}